\documentclass[prb,aps,twocolumn,groupedaddress,showpacs]{revtex4}
\usepackage{graphicx}
\begin{document}

\title{ Radiative corrections to the excitonic molecule state in 
GaAs microcavities }

\author{A. L. Ivanov}

\address{Department of Physics and Astronomy, Cardiff University, 
Queen's Buildings, Cardiff CF24 3YB, United Kingdom}

\author{P. Borri, W. Langbein, and U. Woggon}

\address{Lehrstuhl f\"{u}r Experimentelle Physik
EIIb, Universit\"{a}t Dortmund, Otto-Hahn Str.4, 44227 Dortmund, Germany}

\date{\today}

\begin{abstract}

The optical properties of excitonic molecules (XXs) in GaAs-based 
quantum well microcavities (MCs) are studied, both theoretically and 
experimentally. We show that the radiative corrections to the XX state, 
the Lamb shift $\Delta^{\rm MC}_{\rm XX}$ and radiative width 
$\Gamma^{\rm MC}_{\rm XX}$, are large, about $10-30\,\%$ of the molecule 
binding energy $\epsilon_{\rm XX}$, and definitely cannot be neglected. 
The optics of excitonic molecules is dominated by the in-plane resonant 
dissociation of the molecules into outgoing 1$\lambda$-mode and 
0$\lambda$-mode cavity polaritons. The later decay channel, ``excitonic 
molecule $\rightarrow$ 0$\lambda$-mode polariton +  0$\lambda$-mode 
polariton'',  deals with the short-wavelength MC polaritons invisible in 
standard optical experiments, i.e., refers to ``hidden'' optics of 
microcavities. By using transient four-wave mixing and pump-probe 
spectroscopies, we infer that the radiative width, associated with 
excitonic molecules of the binding energy $\epsilon_{\rm XX} \simeq 
0.9-1.1$\,meV, is $\Gamma^{\rm MC}_{\rm XX} \simeq 0.2-0.3$\,meV in the 
microcavities and $\Gamma^{\rm QW}_{\rm XX} \simeq 0.1$\,meV in a 
reference GaAs single quantum well (QW). We show that for our 
high-quality quasi-two-dimensional nanostructures the $T_2 = 2 T_1$ 
limit, relevant to the XX states, holds at temperatures below 10\,K, 
and that the bipolariton model of excitonic molecules explains 
quantitatively and self-consistently the measured XX radiative widths. 
A nearly factor two difference between $\Gamma^{\rm MC}_{\rm XX}$ and 
$\Gamma^{\rm QW}_{\rm XX}$ is attributed to a larger number of the XX 
optical decay channels in microcavities in comparison with those in 
single QWs. We also find and characterize two critical points in the 
dependence of the radiative corrections against the microcavity detuning, 
and propose to use the critical points for high-precision measurements of 
the molecule binding energy and microcavity Rabi splitting. 

\end{abstract}
\maketitle

\section{Introduction}

The optical properties of an excitonic molecule originate from the 
resonant interaction of its constituent excitons (Xs) with the light 
field. For semiconductor (GaAs) nanostructures we analyze in this paper, 
the above interaction refers to quasi-two-dimensional (quasi-2D) QW 
excitons and is different in single, MC-free quantum wells and in 
microcavities. In the first case, the breaking of translational 
invariance along the growth direction ($z$-direction) leads to the 
coupling of QW excitons to a continuum of bulk photon modes. This 
results in an irreversible radiative decay of low-energy QW excitons 
into the bulk photon modes and to interface, or QW, polaritons for 
the QW exciton states lying outside the photon cone 
\cite{Agranovich66,Nakayama85,Andreani90}. An interface polariton is 
the in-plane propagating eigenwave guided by a single QW, and the light 
field associated with interface polaritons is evanescent, i.e., it 
decays exponentially in the $z$-direction. In contrast, the MC polariton 
optics deals with the quasi-stationary mixed states of quasi-2D MC 
photons and QW excitons \cite{Weisbuch92,Savona94}, i.e., one realizes a 
nearly pure 2D exciton-photon system with resonant coupling between two 
eigenmodes (for a review of the MC polariton optics see, e.g., 
Refs.\,[\onlinecite{SkolnickSST98,KhitrovaRMP99}]). In this case the 
radiative lifetime of MC polaritons originates from a finite transmission 
through the cavity mirrors. The main aim of the present work is to 
develop coherent optics of quasi-2D excitonic molecules in semiconductor 
microcavities. 

The XX-mediated optical response from GaAs microcavities has been 
addressed only recently 
\cite{Gonokami97,Fan98,BorriPRBRb00,Jacob00,Saba00,Baars01,Tartakovskii02}. 
The Coulombic attractive interaction of cross-circular polarized 
($\sigma^+$ and $\sigma^-$) excitons, which gives rise to the XX bound 
state, has been invoked and estimated in order to analyze the 
frequency-degenerate four-wave mixing (FWM) experiment \cite{Gonokami97}. 
Pump-probe specroscopy was used in Ref.\,[\onlinecite{Fan98}] to observe 
the XX-mediated pump-induced changes in the MC reflectivity spectrum. 
However, in the above first experiments the microcavity polariton 
resonance has large broadening so that the spectrally--resolved XX 
transition was not detected. Only recently the spectrally--resolved 
``polariton $\leftrightarrow$ XX'' photon-assisted transition in 
GaAs-based MCs has been observed by using differential reflection 
spectroscopy \cite{BorriPRBRb00,Jacob00}. In particular, the transition 
is revealed in a pump--probe experiment as an induced absorption from 
the lower polariton dispersion branch to the XX state, at positive 
pump-probe time delays \cite{BorriPRBRb00}. In the latter work the MC 
Rabi splitting $\Omega_{\rm 1\lambda}^{\rm MC}$, associated with a 
heavy--hole QW exciton, exceeds the XX binding energy 
$\epsilon_{\rm XX}$ for more than a factor of three. The last experiments 
on excitonic molecules in GaAs microcavities use a high-intensity laser 
field to investigate the XX-mediated changes in the polariton spectrum 
\cite{Saba00,Baars01} and parametric scattering of MC polaritons 
\cite{Tartakovskii02}. In this work we are dealing with a 
{\it low-intensity} limit of the XX optics, aiming to study the radiative 
corrections to the molecule state in a high-quality GaAs single QW 
embedded in a co-planar $\lambda$-cavity. Recently, the optical properties 
of large binding energy excitonic molecules in a MC-embedded ZnSe QW has 
been studied \cite{Neukirch00}. The theoretical model we work out can 
straightforwardly be adapted to the quasi-2D molecules in II-VI 
nanostructures. 

In the previous theoretical studies \cite{LaRoccaJOSAB98,SiehEPJ99} the 
XX radiative corrections are not included, so that the models deal 
with the optically unperturbed molecule wavefunction $\Psi_{\rm XX}$ and 
binding energy $\epsilon_{\rm XX}$. According to 
Ref.\,[\onlinecite{LaRoccaJOSAB98}], the XX radiative corrections are 
rather small, even if $\Omega_{\rm 1\lambda}^{\rm MC} \gg 
\epsilon_{\rm XX}$. The authors argue qualitatively that a volume of 
phase space, where the resonant coupling of the constituent excitons with 
the light field occurs, is rather small to affect the XX state. As we 
show below, an exactly-solvable bipolariton model \cite{Ivanov95,Ivanov98}, 
adapted to excitonic molecules in GaAs-based quasi-2D nanostructures, 
yields $\Gamma^{\rm MC}_{\rm XX}$ and $\Delta^{\rm MC}_{\rm XX}$ of about 
$(0.15-0.30)\,\epsilon_{\rm XX}$ for microcavities, and 
$\Gamma^{\rm QW}_{\rm XX}$ and $\Delta^{\rm QW}_{\rm XX}$ of about 
$(0.10-0.15)\,\epsilon_{\rm XX}$ for single QWs. The calculated values 
refer to the weak confinement of QW excitons and QW excitonic molecules 
we deal with in our study. In the weak confinement limit, the 
QW thickness $d_z$ is comparable with the in-plane radius of the above 
electron-hole bound complexes, which are still constructed in terms of 
well-defined transversly-quantized quasi-2D electronic states. In contrast, 
in the strong confinement limit, $d_z$ is much less than the in-plane 
radius of QW excitons (excitonic molecules). 

The radiative corrections to the XX state cannot be neglected, because 
the exciton-photon coupling (polariton effect) changes the dispersion of 
excitons not only in a very close vicinity of the resonant crossover 
between the relevant exciton and photon energies, but in a rather broad 
band $p \sim p_0$. Here wavevector $p_0$ is given by the resonant 
condition $\hbar \omega^{\gamma}(p_0) = cp / \sqrt{\varepsilon_b} = 
E_{\rm X}(p_0)$ between the bulk photon and exciton dispersions 
($\varepsilon_b$ is the background dielectric constant). For GaAs 
structures $p_0 \simeq 2.7 \times 10^5\,{\rm cm}^{-1}$. The dimensionless 
parameter $\delta^{\rm(D)}_{\rm R}$, which scales the XX radiative 
corrections, is $\delta^{\rm(D)}_{\rm R} = (a^{\rm (D)}_{\rm XX} 
p_0)^{\rm D}$, where $a^{\rm (D)}_{\rm XX}$ is the molecule radius 
and $D$ is the dimensionality of a semiconductor structure. Remarkably, 
as we demonstrate below, $\delta^{\rm(2D)}_{\rm R}$ does not depend upon 
the MC detuning between the $\lambda$-cavity mode and $E_{\rm X}$, i.e., 
is the same for microcavities and single QWs. For our high-quality GaAs 
QWs with weak confinement of excitons one estimates $a^{\rm (2D)}_{\rm XX} 
\simeq 200\,\AA$, so that $\delta^{\rm(2D)}_{\rm R} \simeq 0.3$. The 
latter value clearly shows that the exciton-photon coupling does change 
considerably the quasi-2D XX states. Even for the X wavevectors far away 
from the resonant crossover point $p_0$, the polariton effect can still 
have a considerable impact on the dispersion of optically-dressed 
excitons in bulk semiconductors and QWs. To illustrate this, note that 
for bulk GaAs, e.g., the effective mass associated with the upper 
polariton dispersion branch at $p=0$ is given by $M_{\rm eff} \simeq 
M_x/4$, i.e., by factor four is less than the translational mass $M_x$ 
of optically undressed excitons. In a similar way, the dispersion of QW 
excitons dressed by MC photons, which gives rise to 
$\Delta_{\rm XX}^{\rm MC}$ and $\Gamma^{\rm MC}_{\rm XX}$, refers to the 
in-plane wavevector domain $p_{\|} \lesssim p_0$ rather than to a close 
vicinity of the crossover point $p_{\|} \simeq 0$. 

An excitonic molecule can be described in terms of two quasi-bound 
polaritons (bipolariton), if the coupling of the molecule with the 
light field is much stronger than the incoherent scattering processes. 
In this case the sequence ``two incoming polaritons (or bulk photons) 
$\rightarrow$ quasi-bound XX state $\rightarrow$ two outgoing polaritons'' 
is a completely coherent process of the resonant polariton-polariton 
scattering and can be described in terms of the bipolariton wavefunction 
${\tilde \Psi}_{\rm XX}$. The latter includes an inherent contribution 
from the outgoing (incoming) polaritons and should be found from the 
bipolariton wave equation. The solution also yields the radiative 
corrections to the XX energy, i.e., $-\epsilon_{\rm XX} = - 
\epsilon_{\rm XX}^{(0)} + \Delta_{\rm XX} - i \Gamma_{\rm XX}/2$, where 
$\epsilon_{\rm XX}^{(0)}$ is the ``input'' XX binding energy of an 
optically inactive molecule. For some particular model potentials of 
$\sigma^+$-exciton -- $\sigma^-$-exciton interaction, e.g., for the 
deuteron and Gaussian potentials, the bipolariton wave equation can be 
solved exactly \cite{Ivanov95,Ivanov98}. The bipolariton concept was 
verified in high-precision experiments with low-temperature bulk CuCl 
\cite{Chemla79,Akiyama90,Tokunaga99} and CdS \cite{Mann01}, and was also 
applied successfully to explain the XX-mediated optical response from 
GaAs/AlGaAs multiple QWs \cite{Ivanov97}. The latter experiment dealt 
with quasi-2D XXs in the limit of strong QW confinement. In this case 
the bipolariton model shows that the main channel of the optical decay 
of QW excitonic molecules in MC-free structures is the resonant 
photon-assisted dissociation of the molecule into two outgoing interface 
(QW) polaritons. Note that the Coulombic interaction between two 
constituent excitons of the molecule couples the radiative modes and the 
interface  polariton states, so that an ``umklapp'' process between the 
modes can intrinsically be realized. The above picture refers to the 
following scenario of the coherent optical generation and dissociation of 
QW molecules: ``$\sigma^+$ bulk photon + $\sigma^-$ bulk photon 
$\rightarrow$ $\sigma^+$ virtual QW exciton + $\sigma^-$ virtual QW 
exciton $\rightarrow$ QW molecule $\rightarrow$ $\sigma^+$ interface 
polariton + $\sigma^-$ interface polariton''. 

The experiments we report on deal with weakly confined QW excitonic 
molecules, i.e., the QW thickness $d_z = 250\,\AA$ is comparable with 
the radius of excitons in bulk GaAs. The quasi-2D weak confinement 
allows us to neglect inhomogeneous broadening in the detected X- and 
XX-mediated signals. The MC-free single QW is used as a reference 
structure: All the $\lambda$-microcavities, which we study, are embedded 
with a single QW nearly identical to the reference one. By analyzing the 
coherent dynamics of the XX-mediated signal in spectrally-resolved 
transient FWM, we infer the XX radiative width in the microcavities 
and in the reference single QW, $\Gamma^{\rm MC}_{\rm XX}$ and 
$\Gamma^{\rm QW}_{\rm XX}$, respectively. The measurements yield 
$\Gamma^{\rm MC}_{\rm XX}$ larger than $\Gamma^{\rm QW}_{\rm XX}$ by 
nearly factor two. Furthermore, by using pump-probe spectroscopy we 
also estimate the XX binding energies $\epsilon_{\rm XX}^{\rm MC}$ and 
$\epsilon_{\rm XX}^{\rm QW}$. Our measurements deal with the MC detuning 
band $-2\,\mbox{meV} \lesssim \delta \lesssim +2\,\mbox{meV}$. 

Similarly to quasi-2D XXs in high-quality single QWs, the main mechanism 
of the optical decay of MC molecules is their in-plane resonant 
dissociation into MC polaritons. Thus the coherent optical path of the 
XX-mediated signal in our experiments is given by ``$\sigma^+$ (pump) 
bulk photon + $\sigma^-$ (pump) bulk photon $\rightarrow$ MC molecule 
$\rightarrow$ $\sigma^+$ MC polariton + $\sigma^-$ MC polariton 
$\rightarrow$ $\sigma^+$ (signal) bulk photon + $\sigma^-$ (signal) bulk 
photon''. The latter escape of the MC polaritons into the bulk photon 
modes is due to a finite radiative lifetime of MC photons. In order to 
explain the experimental data, the bipolariton model is adapted to weakly 
confined quasi-2D molecules in (GaAs) microcavities and MC-free (GaAs) 
single QWs. One of the most important features of the optics of excitonic 
molecules in microcavities is a large contribution to the bipolariton 
state ${\tilde \Psi}_{\rm XX}$ from $0\lambda$-mode MC polaritons. The 
relevant $0\lambda$-mode polariton states refer to the in-plane 
wavevectors $p_{\|} \sim p_0$, i.e., are short-wavelength in comparison 
with the $1\lambda$-mode polariton states activated in standard optical 
experiments. An ``invisible'' decay channel of the MC molecule into two 
outgoing $0\lambda$-mode polaritons in combination with the directly 
observable dissociation path ``XX $\rightarrow$ $1\lambda$-mode MC 
polariton + $1\lambda$-mode MC polariton'' explain qualitatively the 
factor two difference between $\Gamma^{\rm MC}_{\rm XX}$ and 
$\Gamma^{\rm QW}_{\rm XX}$. The use of the microcavities embedded with a 
single QW allows us to apply the bipolariton model without complications 
due to the dark X states in multiple QWs \cite{LaRoccaJOSAB98}. The 
bipolariton model quantitatively reproduces our experimental data and 
predicts new spectral features, like $M_{1,2}$ critical points in the 
detuning dependent $\Gamma^{\rm MC}_{\rm XX} = 
\Gamma^{\rm MC}_{\rm XX}(\delta)$ and $\Delta^{\rm MC}_{\rm XX} = 
\Delta^{\rm MC}_{\rm XX}(\delta)$. 

Thus the main results of our study on weakly confined quasi-2D 
molecules in GaAs microcavities are (i) rigorous justification of the 
bipolariton model, (ii) importance of the XX radiative corrections, 
and (iii) existence of the efficient ``hidden'' XX decay channel, 
associated with 0$\lambda$-mode MC polaritons. 

In Sec.\,II, we apply the bipolariton model in order to analyze the XX 
radiative corrections, the XX Lamb shift $\Delta_{\rm XX}$ and XX 
radiative width $\Gamma_{\rm XX}$, relevant to our microcavities and 
reference QW. After a brief discussion of interface and MC polaritons, 
we demonstrated that in GaAs-based quasi-2D structures the XX radiative 
corrections can be as large as $10-30\,\%$ of the (input) XX binding 
energy $\epsilon^{(0)}_{\rm XX}$. It is shown that independently of the 
MC detuning $\delta$ the XX radiative corrections in microcavities and 
(reference) QWs are scaled by the same dimensionless parameter 
$\delta_{\rm R}^{\rm (2D)} = ( a^{\rm (2D)}_{\rm XX} p_0 )^2$, and that 
the main XX optical decay channels in microcavities are ``XX 
$\rightarrow$ 1$\lambda$-mode MC polariton + 1$\lambda$-mode MC 
polariton'' and ``XX $\rightarrow$ 0$\lambda$-mode MC polariton + 
0$\lambda$-mode MC polariton'' against the main decay path in single 
QWs, ``XX $\rightarrow$ interface polariton + interface polariton''. 
We also find and classify two {\it critical points}, $M_1$ and $M_2$, in 
the spectrum of the XX radiative corrections in microcavities, 
$\Gamma^{\rm MC}_{\rm XX} = \Gamma^{\rm MC}_{\rm XX}(\delta)$ and/or 
$\Delta^{\rm MC}_{\rm XX} = \Delta^{\rm MC}_{\rm XX}(\delta)$, and 
propose to use the critical points for high-precision measurements of 
the MC Rabi splitting and the XX binding energy. 

In Sec.\,III, the investigated GaAs-based MC sample and the reference 
GaAs single QW are characterized. We describe the FWM measurements at 
$T=9$\,K, which allow us to estimate the XX dephasing width for the MC 
detuning band $-2\,\mbox{meV} \lesssim \delta \lesssim 2\,\mbox{meV}$, 
${\tilde \Gamma}^{\rm MC}_{\rm XX}(T\!=\!9\,\mbox{K}) \simeq 
0.3-0.4$\,meV, and the pump-probe experiments at $T=5$\,K, which yield 
the bipolariton (XX) binding energy in our microcavities, 
$\epsilon^{\rm MC}_{\rm XX} \simeq 0.9-1.1$\,meV. 

In Sec.\,IV, by analyzing a temperature-dependent contribution to the 
dephasing widths ${\tilde \Gamma}^{\rm MC}_{\rm XX}$ and 
${\tilde \Gamma}^{\rm QW}_{\rm XX}$, which is associated with XX -- 
LA-phonon scattering, we estimate the corresponding XX radiative widths 
in the microcavities and reference QW ($\Gamma^{\rm MC}_{\rm XX} \simeq 
0.2-0.3$\,meV and $\Gamma^{\rm QW}_{\rm XX} \simeq 0.1$\,meV), and show 
that the bipolariton model does reproduce {\it quantitatively and 
self-consistently} both $\Gamma^{\rm MC}_{\rm XX}$ and 
$\Gamma^{\rm QW}_{\rm XX}$. It is shown that the $T_2 = 2 T_1$ limit, 
which is crucial for the validity of the bipolariton model, starts to 
hold for excitonic molecules at cryostat temperatures below 10\,K. We 
also discuss the underlaying physical picture responsible for the large 
XX radiative corrections in high-quality quasi-2D (GaAs) nanostructures. 

A short summary of the results is given in Sec.\,V.

\section{ The bipolariton states in microcavities and single quantum 
wells }

In this Section we briefly discuss interface (quantum well) and 
microcavity polaritons, and apply the bipolariton model 
\cite{Ivanov95,Ivanov98} in order to calculate the XX radiative 
corrections and to describe the optical decay channels of excitonic 
molecules in high-quality GaAs-based microcavities and single QWs. 

\subsection{Interface and microcavity polaritons}

For a single QW, the resonant coupling of excitons with the light field 
can be interpreted in terms of the radiative in-plane modes 
$|{\bf p}_{\|}| \leq p_0$, which ensure communication of low-energy QW 
excitons with incoming and outgoing bulk photons (the only photons used 
in standard pump-probe optical experiments with QWs), and interface 
polaritons, which refer to the states outside the photon cone, 
$|{\bf p}_{\|}| \geq \omega_t \sqrt{\varepsilon_b}/c$. The latter 
in-plane propagating polariton eigenmodes are trapped and waveguided by 
the X resonance; they are accompanied by the evanescent, interface light 
field, i.e., are invisible at macroscopic distances from the QW. 

For an ideal QW microcavity the MC photons with in-plane wavevector 
${\bf p}_{\|}$ can be classified in terms of $n \lambda$- transverse 
eigenmodes ($n=0,1,2,...$). The MC polariton eigenstates arise when some 
of the MC photon eigenmodes resonate with the QW exciton state. As we 
show below, only $0 \lambda$- and $1 \lambda$- polariton eigenmodes are 
relevant to the optics of QW excitonic molecules in our MC structures. 
With increasing MC thickness towards infinity the microcavity polariton 
eigenstates evolve into the radiative and interface polariton eigenmodes 
associated with a MC-free single QW \cite{Savona94}. 

(i) {\it The light field resonantly interacting with quasi-2D 
excitons in a single (GaAs) QW.}
The interaction of a QW exciton with in-plane momentum $\hbar {\bf p}_{\|}$ 
with the transverse light field of frequency $\omega$ is characterized by 
the dispersion equation \cite{Agranovich66,Nakayama85,Andreani90}: 
\begin{equation}
{ c^2 p_{\|}^2 \over \varepsilon_b } = 
\omega^2 + { \omega^2 R^{\rm QW}_{\rm X} \sqrt{p_{\|}^2 - \varepsilon_b 
(\omega/c)^2} \over \omega^2_t + \hbar \omega_t p_{\|}^2/M_x 
- i \omega \gamma_{\rm X} - \omega^2 \ } \ ,  
\label{pol}
\end{equation}
where $M_x$ is the in-plane translational X mass, $\hbar \omega_t = 
E_{\rm X}({\bf p}_{\|}$=0) is the X energy, $\gamma_{\rm X}$ is the 
rate of incoherent scattering of QW excitons, and $R^{\rm QW}_{\rm X}$ 
is the dimensional oscillator strength of exciton-photon interaction 
per QW unit area. Equation (\ref{pol}) refers to a single (GaAs) 
QW confined by two identical (AlGaAs) bulk barriers. 

For $|{\bf p}_{\|}| \geq \omega \sqrt{\varepsilon_b}/c$, i.e., for the 
momentum-frequency domain outside the photon cone, Eq.~(\ref{pol}) 
describes the in-plane polarized transverse interface polaritons ($Y$-mode 
polaritons). The evanescent light field associated with the interface 
polaritons is given by ${\bf E}(\omega,{\bf p}_{\|},z) = 
{\bf E}(\omega,{\bf p}_{\|}) \exp(-\kappa |z|)$, where 
$\kappa = \sqrt{p_{\|}^2 - \varepsilon_b (\omega/c)^2}$. The exciton 
and photon components of a QW polariton with in-plane wavevector 
${\bf p}_{\|}$ are 
\begin{eqnarray}
&& u^2_{\rm IP}(p_{\|}) = { \kappa R^{\rm QW}_{\rm X} \over \kappa 
R^{\rm QW}_{\rm X} + 2 [ \omega_t  + \hbar  p_{\|}^2/2M_x -  
\omega_{\rm IP}(p_{\|}) ]^2 } \ , 
\nonumber \\
&& v^2_{\rm IP}(p_{\|}) = 1 - u^2_{\rm IP}(p_{\|}) \ , 
\label{comp}
\end{eqnarray}
respectively. Here $\omega = \omega_{\rm IP}(p_{\|})$ is the polariton 
dispersion determined by Eq.\,(\ref{pol}). Note that the $z$-polarized 
transverse interface polaritons ($Z$-mode QW polaritons) associated with 
the ground-state heavy-hole excitons are not allowed in GaAs QWs 
\cite{Andreani90}. 

The low-energy QW excitons from the radiative zone $|{\bf p}_{\|}| \leq 
p_0 = \omega_t \sqrt{\varepsilon_b}/c$ couple with bulk photons, i.e., 
can radiatively decay into the bulk photon modes. In this case 
Eq.\,(\ref{pol}) yields the X radiative decay rate into bulk in-plane 
($Y$-) polarized transverse photons:  
\begin{equation} 
{1 \over \hbar}\Gamma^{\rm QW}_{\rm X}(p_{\|}) = {\varepsilon_b 
\over c^2} R^{\rm QW}_{\rm X} { \omega_t \over \sqrt{p^2_0 - 
p_{\|}^2} } \ .
\label{rad} 
\end{equation} 
One can also re-write Eq.\,(\ref{rad}) as 
$\Gamma^{\rm QW}_{\rm X}(p_{\|}) = \Gamma^{\rm QW}_{\rm X}(p_{\|}$=0) 
$p_0 / (p^2_0 - p_{\|}^2)^{1/2}$, where 
$\Gamma^{\rm QW}_{\rm X}(p_{\|}$=0) = $\hbar (\sqrt{\varepsilon_b} / c) 
R^{\rm QW}_{\rm X}$ is the radiative width of a QW exciton with 
in-plane momentum $\hbar p_{\|}=0$. In high-quality GaAs QWs 
at low temperatures, the condition $\Gamma^{\rm QW}_{\rm X} \gg \hbar 
\gamma_{\rm X}$ can be achieved, so that the X dispersion within the 
photon cone is approximated by $\hbar 
\omega^{\rm QW}_{\rm X}(p_{\|}$$\leq$$p_0) = \hbar \omega_t + \hbar^2 
p_{\|}^2/2M_x - i \Gamma^{\rm QW}_{\rm X}(p_{\|})/2$. 

The oscillator strength $R_{\rm X}^{\rm QW}$ associated with QW excitons 
is given by 
\begin{equation}
R_{\rm X}^{\rm QW} = {4 \pi \over \hbar} \, {\omega_t \over \varepsilon_b} 
\, |\phi_{\rm X}^{\rm (2D)}({\bf r} = 0)|^2 \, |d_{\rm cv}|^2 \, ,
\label{QWstr}
\end{equation}
where $\phi_{\rm X}^{\rm (2D)}(r)$ is the X wavefunction of relative 
electron-hole motion, and $d_{\rm cv}$ is the dipole matrix element of 
the interband optical transition. In the limits of strong and weak QW 
confinement Eq.\,(\ref{QWstr}) yields 
\begin{equation}
R^{\rm QW}_{\rm X} = \left\{
\begin{array}{ll}
16 a_{\rm X}^{\rm (3D)} \omega_{\ell t} \omega_t \, , 
\ \ \ \ \ \  \ \ \ \ \ \ \ \ a^{\rm (3D)}_{\rm X} \gg d_z \, , \\
2d_z \omega_{\ell t} \omega_t \, , \ \ \ \ \ \lambda = 2 \pi/p_0 \gg  
d_z \gtrsim a^{\rm (3D)}_{\rm X} \, , \\ 
\end{array}
\right.
\label{strength}
\end{equation}
respectively, where $a^{\rm (3D)}_{\rm X}$ is the Bohr radius of bulk 
excitons and $\omega_{\ell t}$ is the longitudinal-transverse splitting 
associated with bulk excitons (in bulk GaAs one has $a^{\rm (3D)}_{\rm X} 
\simeq 136\,\AA$ and $\hbar \omega_{\ell t} \simeq 80-86\,\mu{\rm eV}$, 
respectively \cite{Ulbrich82}). Thus we estimate the upper limit of the 
oscillator strength in narrow GaAs QWs as $\hbar^2 
R^{\rm QW}_{\rm X}(d_z$$\rightarrow$$0) \simeq 0.26 - 0.28\,{\rm eV}^2\AA$. 
For our GaAs QWs with weak confinement of excitons one evaluates from 
Eq.\,(\ref{strength}) that $\hbar^2 R^{\rm QW}_{\rm X}(d_z$$=$$250\,\AA) 
\simeq 0.061\,{\rm eV}^2\AA$. 

(ii) {\it The MC polariton dispersion relevant to excitonic molecules 
in (GaAs-based) microcavities.} 
The dispersion equation for MC polaritons, which contribute to the 
XX-mediated optics of a $\lambda$-cavity we study in our experiments, is 
given by 
\begin{eqnarray}
\omega^2_t &+& 
\hbar \omega_t p_{\|}^2/M_x - i \omega \gamma_{\rm X} - \omega^2 = 
\nonumber \\ 
&=&  \omega^2 \Bigg[ { (\Omega^{\rm MC}_{\rm 1 \lambda})^2 \over 
(\omega^{\gamma}_{1 \lambda})^2 - i \omega \gamma_{\rm R} - \omega^2 } 
\ + \ { (\Omega^{\rm MC}_{\rm 0 \lambda})^2 \over 
(\omega^{\gamma}_{0 \lambda})^2 - i \omega \gamma_{\rm R} - 
\omega^2 } \Bigg] \, , 
\nonumber \\
\label{MC}
\end{eqnarray}
where the photon frequencies, associated with the 1$\lambda$- and 
0$\lambda$- microcavity eigenmodes, are $\omega^{\gamma}_{1 \lambda} =  
\omega^{\gamma}_{1 \lambda}(p_{\|}) = ( c^2 p_{\|}^2/\varepsilon_b + 
\omega_0^2 )^{1/2}$ and $\omega^{\gamma}_{0 \lambda} =  
\omega^{\gamma}_{0 \lambda}(p_{\|}) = c p_{\|}/\sqrt{\varepsilon_b}$, 
respectively. Here $\omega_0 = (2 \pi c)/(L_z \sqrt{\varepsilon_b})$ is 
the cavity eigenfrequency, $L_z$ is the MC thickness, and 
$\gamma_{\rm R}$ is the inverse radiative lifetime of MC photons, due 
to their escape from the microcavity into external bulk photon modes. 
The MC Rabi frequency $\Omega^{\rm MC}_{\rm 1 \lambda}$ refers to 
1$\lambda$-eigenmode of the light field, ${\hat {\rm e}}_{1 \lambda}(z) 
= \sqrt{2/L_z} \cos[(2 \pi z)/L_z]$ (we assume that the QW is located 
at $z=0$ so that $|z| \leq L_z/2$), and is determined by 
\begin{equation}
(\Omega^{\rm MC}_{\rm 1 \lambda})^2 = {16 \pi \over \hbar} \, 
{\omega_t \over \varepsilon_b} \, |\phi_{\rm X}^{\rm (2D)}({\bf r} 
= 0)|^2 \, |d_{\rm cv}|^2 \, { |I_1|^2 \over L_z} \, ,
\label{MCstr1}
\end{equation}
where $I_1 = I_1(d_z/L_z) = [L_z/(\pi d_z)] \sin[(\pi d_z)/L_z] \simeq 1 - 
(\pi/6)(d_z/L_z)^2$. In turn, the Rabi frequency $\Omega^{\rm MC}_{\rm 0 
\lambda}$ is associated with 0$\lambda$-eigenmode of the MC light field, 
${\hat {\rm e}}_{0 \lambda}(z) = 1/\sqrt{L_z} = {\rm const.}$, and 
\begin{equation}
(\Omega^{\rm MC}_{\rm 0 \lambda})^2 = {8 \pi \over \hbar} \, 
{\omega_t \over \varepsilon_b} \, |\phi_{\rm X}^{\rm (2D)}({\bf r} = 0)|^2 
\, |d_{\rm cv}|^2 \, { 1 \over L_z} \, .
\label{MCstr0}
\end{equation}
From Eqs.\,(\ref{QWstr}), (\ref{MCstr1}), and (\ref{MCstr0}) one gets
\begin{equation}
(\Omega^{\rm MC}_{\rm 1 \lambda})^2 = 2 |I_1|^2  
(\Omega^{\rm MC}_{\rm 0 \lambda})^2 = 4 \, { R_{\rm X}^{\rm MC} \over 
L_z } \, |I_1|^2  \, .
\label{strR}
\end{equation}
Because the factor $|I_1|^2 \simeq 1$ (for our microcavities $d_z = 
250\,\AA$ and $L_z \simeq 2326\,\AA$, so that $|I_1|^2 \simeq 0.96$), 
we conclude that $\Omega^{\rm MC}_{\rm 1 \lambda} \simeq \sqrt{2} 
\Omega^{\rm MC}_{\rm 0 \lambda} \simeq 2 (R_{\rm X}^{\rm MC}/L_z)^{1/2}$. 
The factor two difference between $(\Omega^{\rm MC}_{\rm 1 \lambda})^2$ 
and $(\Omega^{\rm MC}_{\rm 0 \lambda})^2$ originates from the difference 
of the intensities of the light fields associated with microcavity 
1$\lambda$- and 0$\lambda$-eigenmodes at the QW position, $z=0$, i.e., 
is due to $|{\hat {\rm e}}_{1 \lambda}(z$=$0)|^2/|{\hat {\rm e}}_{0 
\lambda}(z$=$0)|^2 = 2$. 

\begin{figure}
\includegraphics*[width=8cm]{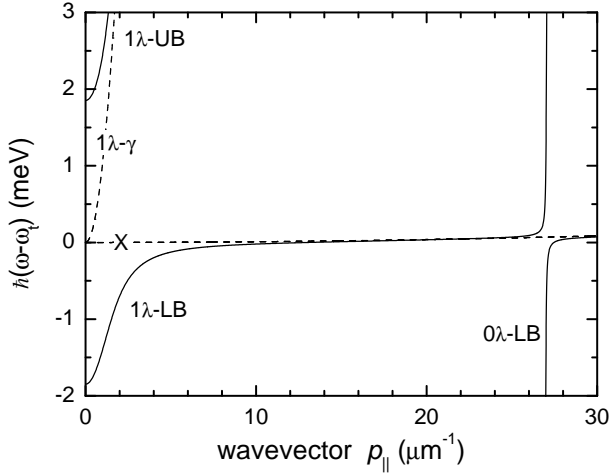}
\caption[]{ Three-branch microcavity polariton dispersion calculated 
with Eq.\,(\ref{MC}) for zero-detuning. The parameters are adapted to 
the GaAs microcavities used in our experiments: $\hbar 
\Omega^{\rm MC}_{1\lambda} = 3.70$\,meV, $\hbar \Omega^{\rm MC}_{0\lambda} 
= 2.67$\,meV, $\varepsilon_b = 12.3$, $M_x = 0.4\,m_0$, and 
$E_{\rm X}(p_{\|}$=$0) = \hbar \omega_t = 1.5219$\,eV. The dashed lines 
show the 1$\lambda$-mode MC photon and exciton dispersions (the 
0$\lambda$-mode photon dispersion is not plotted). } 
\end{figure}

Thus, the dispersion Eq.\,(\ref{MC}) deals with a {\it three-branch} 
MC polariton model. In Fig.\,1 we plot the polariton dispersion 
branches, designated by 1$\lambda$-UB (upper branch), 1$\lambda$-LB 
(middle branch), and 0$\lambda$-LB (lower branch), respectively, and 
calculated by Eq.\,(\ref{MC}) for a zero-detuning GaAs-based microcavity 
with $\hbar \Omega^{\rm MC}_{\rm 1 \lambda} = 3.70$\,meV and $\hbar 
\Omega^{\rm MC}_{\rm 0 \lambda}= 2.67$\,meV. The ratio between the Rabi 
frequencies satisfies Eq.\,(\ref{strR}), and the used value of 
$\Omega^{\rm MC}_{\rm 1 \lambda}$ corresponds to that observed in our 
experiments. For small in-plane wavevectors $|{\bf p}_{\|}| \lesssim 
p_{\|}^{(1 \lambda)} \simeq 0.5 \times 10^5\,{\rm cm}^{-1}$ (see 
Fig.\,1) the 1$\lambda$-UB and 1$\lambda$-LB dispersion curves are 
identical to the upper and lower MC polariton branches calculated within 
the standard 1$\lambda$-eigenmode resonant approximation 
\cite{SkolnickSST98,KhitrovaRMP99}. In this case the 1$\lambda$-UB 
and 1$\lambda$-LB polaritons are purely 1$\lambda$-eigenwaves; the 
0$\lambda$-LB dispersion is well-separated from the X resonance so that 
in Eq.\,(\ref{MC}) one can put $\Omega^{\rm MC}_{\rm 0 \lambda} = 0$ in 
order to describe the 1$\lambda$-UB and 1$\lambda$-LB dispersions in the 
wavevector domain $p_{\|} \lesssim p_{\|}^{(1 \lambda)}$. The 
anti-crossing between the X dispersion $\omega_t + \hbar p_{\|}^2/2M_x$ 
and the MC 0$\lambda$-mode photon frequency $c p_{\|}/
\sqrt{\varepsilon_b}$, which occurs at $p_{\|} = p_0 \simeq 2.7 \times 
10^5\,{\rm cm}^{-1}$, gives rise to the MC 0$\lambda$-eigenmode 
dispersion associated with the 1$\lambda$-LB and 0$\lambda$-LB 
short-wavelength polaritons with $p_{\|} \gg p_{\|}^{(1\lambda)}$  
(see Fig.\,1). This picture is akin to the two-branch polariton 
dispersion in bulk semiconductors; for $p_{\|} \simeq p_0$ the 
1$\lambda$-LB and 0$\lambda$-LB polariton dispersion can accurately 
be approximated by Eq.\,(\ref{MC}) with $\Omega^{\rm MC}_{\rm 1 \lambda} 
= 0$. In this case  Eq.\,(\ref{MC}) becomes identical to the dispersion 
equation for bulk polaritons, if in the latter the bulk Rabi splitting 
$\Omega^{\rm bulk}$ ($\hbar \Omega^{\rm bulk} \simeq 15.6$\,meV in GaAs) 
is replaced by $\Omega^{\rm MC}_{\rm 0 \lambda}$ and the bulk photon 
wavevector $p$ is replaced by $p_{\|}$. Note that for the MC 
0$\lambda$-eigenmode the light field is homogeneous in the $z$-direction 
within the microcavity, i.e., for $|z| \leq L_z/2$. With increasing 
detuning from the X resonance the 1$\lambda$-LB and 0$\lambda$-LB 
polariton dispersions approach the photon frequencies 
$\omega^{\gamma}_{0 \lambda} = c p_{\|} / \sqrt{\varepsilon_b}$ and 
${\tilde \omega}^{\gamma}_{0 \lambda} = c p_{\|} / 
\sqrt{\varepsilon^{(0)}_b}$, respectively, where the low-frequency 
dielectric constant is given by $\varepsilon^{(0)}_b = \varepsilon_b [1 + 
(\Omega_{0\lambda}^{\rm MC}/\omega_t)^2]$. The interconnection between 
two MC polariton domains occurs via the 1$\lambda$-LB polariton 
dispersion: With increasing $p_{\|}$ from $p_{\|} \lesssim p_{\|}^{(1 
\lambda)}$ towards $p_{\|} \gtrsim p_0$ the structure of the photon 
component of 1$\lambda$-LB polaritons smoothly changes, as a 
superposition of two modes, from purely 1$\lambda$-mode to purely 
0$\lambda$-mode. 


Because $1/a^{\rm (2D)}_{\rm XX} > p_0$, the non-zero exciton component 
of all three MC polariton dispersion branches contributes to the molecule 
state and, therefore, to the XX-mediated optics of microcavities. The 
X component, associated with the 0$\lambda$-LB, 1$\lambda$-LB, and 
1$\lambda$-UB dispersions, is given by 
\begin{eqnarray}
(u^{\rm MC}_i)^2 = \Bigg[ 1 &+& { \omega^4_i (\Omega_{1 
\lambda}^{\rm MC})^2 \over \omega_t \omega^{\gamma}_{1 \lambda}
[\omega^2_i - (\omega^{\gamma}_{1 \lambda})^2]^2 } 
\nonumber \\
&+& { \omega^4_i (\Omega_{0 \lambda}^{\rm MC})^2 \over \omega_t 
\omega^{\gamma}_{0 \lambda} [\omega^2_i - 
(\omega^{\gamma}_{0 \lambda})^2]^2 } \Bigg]^{-1} \, , 
\label{MCcomp}
\end{eqnarray}
where $\omega_{i = {\rm 0\lambda LB,1\lambda LB, 1\lambda UB}} = 
\omega_{i = {\rm 0\lambda LB,1\lambda LB, 1\lambda UB}}^{\rm MC}(p_{\|})$ 
are the polariton dispersion branches calculated with Eq.\,(\ref{MC}). 
For a given $p_{\|}$ the X components satisfy the sum rule, 
$(u^{\rm MC}_{\rm 0\lambda LB})^2 + (u^{\rm MC}_{\rm 1\lambda LB})^2 + 
(u^{\rm MC}_{\rm 1\lambda UB})^2 = 1$. The exciton components, which 
correspond to the 0$\lambda$-LB, 1$\lambda$-LB, and 1$\lambda$-UB 
dispersions shown in Fig.\,1, are plotted in Fig.\,2. The above 
polariton branches have non-zero X component when the frequencies 
$\omega_i^{\rm MC}(p_{\|})$ resonate with the X state, i.e., at $p_{\|} 
\lesssim p_{\|}^{(1 \lambda)}$ for the 1$\lambda$-UB, at $p_{\|} \lesssim 
p_0$ for the 1$\lambda$-LB, and at $p_{\|} \gtrsim p_0$ for the 
0$\lambda$-LB, respectively (see Fig.\,2). In our microcavities the X 
component of the $2\lambda$-, $3\lambda$- etc. eigenmode MC polaritons 
is negligible. 

\begin{figure}
\includegraphics*[width=8cm]{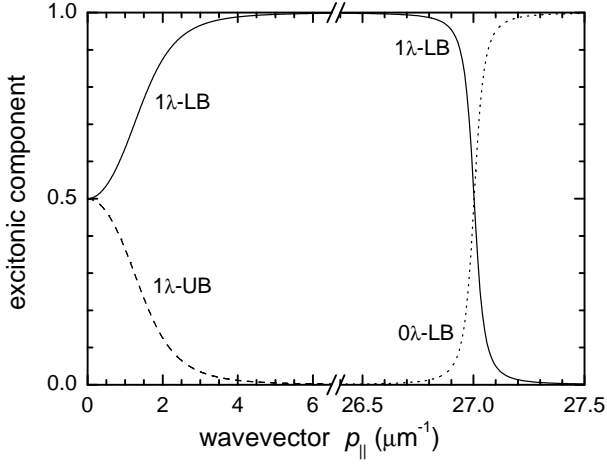}
\caption[]{ The exciton component of 0$\lambda$-LB (dotted line), 
1$\lambda$-LB (solid line), and 1$\lambda$-UB (dashed line) polaritons 
in a zero-detuning GaAs microcavity. } 
\end{figure}

A non-ideal optical confinement of the MC photon modes by distributed 
Bragg reflectors (DBRs) leads to the leakage of MC photons and gives rise 
to the radiative rate $\gamma_{\rm R}$ in Eq.\,(\ref{MC}). Thus the 
radiative width of MC polaritons, due to their optical escape through 
the DBRs, is $\Gamma^{\rm MC}_{i={\rm 0\lambda LB,1\lambda LB,1\lambda 
UB}} = \hbar (v^{\rm MC}_{i={\rm 0\lambda LB,1\lambda LB,1\lambda UB}})^2 
\gamma_{\rm R}$, where the photon component of the polaritons is given 
by $(v_i^{\rm MC})^2 = 1 - (u_i^{\rm MC})^2$. Note that for our 
GaAs-based macrocavities at low temperatures, $\gamma_{\rm R}$ is much 
larger than $\gamma_{\rm X}$ \cite{LangbeinPRB2000,BorriPRB2000}.

\subsection{Bipolaritons in GaAs quantum wells}

The quasi-2D excitonic molecules in single QWs without co-planar optical 
confinement of the light field can either resonantly dissociate into 
interface polaritons or decay radiatively into the bulk photon modes. In our 
optical experiments, which deal with pump and signal bulk photons only, 
the first route of the XX optical decay cannot be visualized directly. Thus 
this channel refers to the ``hidden'' optics associated with the evanescent 
light field resonantly guided by QW excitons.

(i) {\it Resonant dissociation of QW excitonic molecules into outgoing 
interface polaritons.} The bipolariton model allows us to calculate the 
XX radiative width $\Gamma_{\rm XX}^{\rm QW(1)} = 
\Gamma_{\rm XX}^{\rm QW(1)}({\bf K}_{\|})$, associated with the resonant 
dissociation of the molecule with in-plane translational momentum 
$\hbar {\bf K}_{\|}$, by solving the wave equation 
\cite{Ivanov95,Ivanov98}: 
\begin{eqnarray}
&&\Big[ E_{\rm IP}({\bf p}_{\|}+{\bf K}_{\|} / 2) +  
E_{\rm IP}(-{\bf p}_{\|} + {\bf K}_{\|} / 2 ) 
\Big] \tilde{\Psi}_{\rm XX}({\bf p}_{\|},{\bf K}_{\|})
\nonumber \\ 
&&+ \ f_{\rm IP}({\bf p}_{\|},{\bf K}_{\|}) \sum_{\bf p'_{\|}} 
W_{\sigma^+\sigma^-}({\bf p}_{\|} - {\bf p'}_{\|}) 
\tilde{\Psi}_{\rm XX}({\bf p'}_{\|},{\bf K}_{\|})
\nonumber \\ 
&& \ \ \ \ \ \ \ \ \ = \tilde{E}_{\rm XX}^{\rm QW}({\bf K}_{\|}) 
\tilde{\Psi}_{\rm XX}({\bf p}_{\|},{\bf K}_{\|}) \, . 
\label{BPint}
\end{eqnarray}
Here $\tilde{\Psi}_{\rm XX}$ and $\tilde{E}^{\rm QW}_{\rm XX}$ are 
the bipolariton (XX) wavefunction and energy, respectively, $E_{\rm IP} = 
\hbar \omega_{\rm IP}$ is the QW polariton energy determined by the 
dispersion Eq.\,(\ref{pol}), $f_{\rm IP}({\bf p}_{\|},{\bf K}_{\|}) = 
u_{\rm IP}^2({\bf p}_{\|}+{\bf K}_{\|} / 2) u_{\rm IP}^2(-{\bf p}_{\|}
+{\bf K}_{\|} / 2)$, where $u_{\rm IP}^2$ is given by Eq.\,(\ref{comp}), 
$\hbar {\bf p}_{\|}$ is the in-plane momentum of the relative motion 
of the optically-dressed constituent excitons, and $W_{\sigma^+\sigma^-}$ 
is the attractive Coulombic potential between $\sigma^+$- and $\sigma^-$- 
polarized QW excitons. The complex bipolariton energy can also be 
rewritten as $\tilde{E}_{\rm XX}^{\rm QW} = 2 E_{\rm X} - 
\epsilon_{\rm XX}^{(0)} + \Delta_{\rm XX}^{\rm QW} - i 
\Gamma_{\rm XX}^{\rm QW}/2$, where $\epsilon_{\rm XX}^{(0)}$ is the XX 
binding energy with no renormalization by the coupling with the vacuum 
light field. For the non-local deuteron model potential 
$W_{\sigma^+\sigma^-}(|{\bf p}_{\|}-{\bf p'}_{\|}|)$, which yields 
within the standard Schr\"odinger two-particle (two-X) equation the 
wavefunction $\Psi_{\rm XX}^{(0)}(p_{\|}) = 2 \sqrt{2 \pi} 
a_{\rm XX}^{\rm (2D)} / [(p_{\|} a_{\rm XX}^{\rm (2D)})^2 + 1]^{3/2}$ 
for an optically inactive molecule, the bipolariton wave 
Eq.\,(\ref{BPint}) is exactly-solvable \cite{Ivanov95}. The input 
parameters of the model are the binding energy $\epsilon_{\rm XX}^{(0)}$ 
and the oscillator strength $R_{\rm X}^{\rm QW}$. Thus the 
exactly-solvable bipolariton model simplifies the exciton-exciton 
interaction, but treats rigorously the (interface) polariton effect. 

(ii) {\it Resonant decay of QW excitonic molecules into the bulk photon 
modes.} The decay occurs when at least one of the constituent excitons 
of a QW molecule moves within the radiative zone, i.e., when 
$|+{\bf p}_{\|} + {\bf K}_{\|}/2| \leq p_0$ and/or $|-{\bf p}_{\|} + 
{\bf K}_{\|}/2| \leq p_0$. Note that the exciton-exciton resonant coherent 
Coulombic scattering within the molecule state intrinsically couples the 
X radiative and QW polariton modes. Thus the XX width, associated with 
the optical decay into the bulk photon modes, is given by 
\begin{eqnarray}
&&\Gamma_{\rm XX}^{\rm QW(2)}({\bf K}_{\|}\!=\!0) = {1 \over \pi}
 \int_0^{p_0} |\Psi_{\rm XX}^{(0)}(2p_{\|})|^2  
\Gamma_{\rm X}^{\rm QW}(p_{\|}) p_{\|} d p_{\|}
\nonumber \\  
&& \ \ \ \ \ \ \ \ 
= \hbar { \sqrt{\varepsilon_b} \over c } R_{\rm X}^{\rm QW} 
{ \sqrt{\chi} \over 4 (1 + \chi)^{5/2} } 
\Big[ (5 + 2 \chi) \sqrt{\chi (1 + \chi)}
\nonumber \\ 
&& \ \ \ \ \ \ \ \ 
+ 3 \ln( \sqrt{1 + \chi} + \sqrt{\chi}) \Big] \, , 
\label{rada}
\end{eqnarray}
where $\chi = 4 \delta^{\rm (2D)}_{\rm R} \equiv 4 (a^{\rm (2D)}_{\rm XX} 
p_0)^2$ and $\Gamma_{\rm X}^{\rm QW}(p_{\|})$ is given by 
Eq.\,(\ref{rad}). In the above integral over the QW radiative zone we 
approximate $\Psi_{\rm XX}^{(0)}$ by the  deuteron wavefunction. For 
$\chi \ll 1$ Eq.\,(\ref{rada}) yields $\Gamma_{\rm XX}^{\rm QW(2)} \simeq 
2 \hbar (\sqrt{\varepsilon_b}/c) \chi R_{\rm X}^{\rm QW}$ = 
$8 (a^{\rm (2D)}_{\rm XX} p_0)^2 \Gamma_{\rm X}^{\rm QW}(p_{\|}$=0). 
However, for our reference GaAs QW with weak confinement of the electronic 
states one has $\chi \simeq 1.2$ so that the above simple approximation 
of Eq.\,(\ref{rada}) cannot be used. 

\begin{figure}
\includegraphics*[width=8cm]{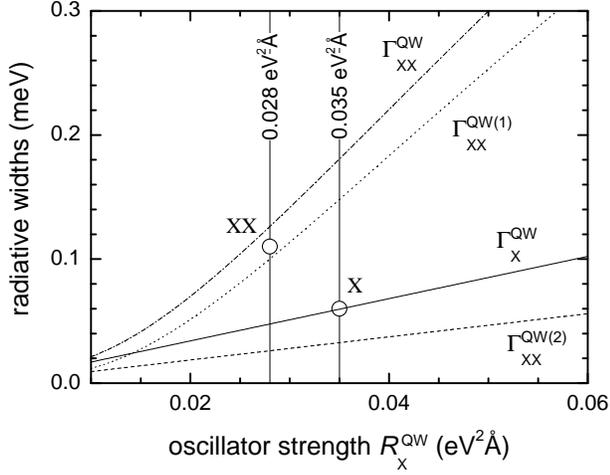}
\caption[]{ The calculated radiative decay widths of the exciton and 
bipolariton states versus the oscillator strength $R^{\rm QW}_{\rm X}$. 
The XX radiative widths associated with the decay into interface 
polaritons, $\Gamma_{\rm XX}^{\rm QW(1)}$, and into bulk photon modes, 
$\Gamma_{\rm XX}^{\rm QW(2)}$, are plotted separately. The input XX 
binding energy $\epsilon^{(0)}_{\rm XX} = 1.1$\,meV. The two circle 
symbols show $\Gamma^{\rm QW}_{\rm X}$ and $\Gamma^{\rm QW}_{\rm XX}$ 
inferred from the experimental data. } 
\end{figure}

In Fig.\,3 we plot the radiative widths 
$\Gamma_{\rm X}^{\rm QW}({\bf p}_{\|}$=0), 
$\Gamma_{\rm XX}^{\rm QW(1)}({\bf K}_{\|}$=0), 
$\Gamma_{\rm XX}^{\rm QW(2)}({\bf K}_{\|}$=0), and 
$\Gamma_{\rm XX}^{\rm QW(1)}({\bf K}_{\|}$=0) + 
$\Gamma_{\rm XX}^{\rm QW(2)}({\bf K}_{\|}$=0) against the oscillator 
strength of QW excitons $R_{\rm X}^{\rm QW}$. The widths are calculated 
with Eqs.\,(\ref{BPint}) and (\ref{rada}) for the input XX binding energy 
$\epsilon_{\rm XX}^{(0)} = 1.1$\,meV. As we discuss in Section III, the 
oscillator strength $R_{\rm X}^{\rm QW}$ of the high-quality reference QW 
used in our experiments is given by 
$\hbar^2 R_{\rm X}^{\rm QW}(d_z$=$250\,\AA) \simeq 0.035\,{\rm eV}^2\AA$. 
The above value, which is inferred from the experimental data, refers to 
the GaAs QW sandwiched between semi-infinite bulk AlGaAs barriers and is 
consistent with that estimated in the previous Subsection by using 
Eq.\,(\ref{strength}). A cap layer on top of the reference single QW 
modifies the evanescent field associated with interface 
polaritons and reduces the oscillator strength to $\hbar^2 
{\tilde R}_{\rm X}^{\rm QW}(d_z$=$250\,\AA) \simeq  0.028\,{\rm eV}^2\AA$  
(for the details see Section IV). As shown in Fig.\,3, for $\hbar^2 
R_{\rm X}^{\rm QW} = 0.035\,{\rm eV}^2\AA$ Eqs.\,(\ref{BPint}) and 
(\ref{rada}) yield $\Gamma_{\rm XX}^{\rm QW(1)}({\bf K}_{\|}$=0) $\simeq 
148\,\mu {\rm eV}$ and $\Gamma_{\rm XX}^{\rm QW(2)}({\bf K}_{\|}$=0) 
$\simeq 33\,\mu {\rm eV}$, so that the total XX radiative width is given by 
$\Gamma_{\rm XX}^{\rm QW}({\bf K}_{\|}$=0) = $\Gamma_{\rm XX}^{\rm QW(1)} 
+ \Gamma_{\rm XX}^{\rm QW(2)} \simeq 0.18$\,meV. For $\hbar^2 
{\tilde R}_{\rm X}^{\rm QW} \simeq  0.028\,{\rm eV}^2\AA$ one calculates 
$\Gamma_{\rm XX}^{\rm QW(1)}({\bf K}_{\|}$=0) $\simeq 100\,\mu {\rm eV}$,   
$\Gamma_{\rm XX}^{\rm QW(2)}({\bf K}_{\|}$=0) $\simeq 26\,\mu {\rm eV}$, 
and $\Gamma_{\rm XX}^{\rm QW}({\bf K}_{\|}$=0) = 
$\Gamma_{\rm XX}^{\rm QW(1)} + \Gamma_{\rm XX}^{\rm QW(2)} \simeq 
0.126$\,meV. The latter value is indeed very close to the XX radiative 
width $\Gamma_{\rm XX}^{\rm QW} \simeq 0.1$\,meV inferred from our 
opical experiments with the reference QW (see Section III). 

The photon-assisted resonant dissociation of QW molecules into outgoing 
interface polaritons is more efficient than the XX optical decay into the 
bulk photon modes by factor 4.5 for $\hbar^2 R_{\rm X}^{\rm QW} \simeq 
0.035\,{\rm eV}^2\AA$ and by factor 3.8 for $\hbar^2 
{\tilde R}_{\rm X}^{\rm QW} \simeq  0.028\,{\rm eV}^2\AA$, respectively. 
This conclusion is consistent with that of Ref.\,[\onlinecite{Ivanov97}], 
where for the limit of strong QW confinement ($d_z \rightarrow 0$) the 
relative efficiency of the two optical decay channels was estimated to be 
$\Gamma_{\rm XX}^{\rm QW(1)} : \Gamma_{\rm XX}^{\rm QW(2)} \simeq 25 : 1$. 
The latter ratio refers to the idealized case of an extremely narrow GaAs 
QW surrounded by infinitely thick AlGaAs barriers. The resonant optical 
dissociation of the QW molecules into interface polaritons is much 
stronger than the radiative decay into the bulk photon modes, because the 
constituent excitons in their relative motion move mainly outside the 
radiative zone, with the in-plane momenta $|\pm {\bf p}_{\|} + 
{\bf K}_{\|}/2| \gtrsim p_0$. In this case the excitons are optically 
dressed by the evanescent light field, i.e., they exist as QW polaritons 
and, therefore, decay mainly into the confined, QW-guided interface 
modes. The picture can also be justified by analyzing the joint density 
of states relevant to the two optical decay channels. Note that in both 
main equations, Eq.\,(\ref{BPint}) and Eq.\,(\ref{rada}), 
$\delta^{\rm (2D)}_{\rm R} = (a^{\rm (2D)}_{\rm XX} p_0)^2$ does 
represent the dimensionless smallness parameter of the (bipolariton) 
model.

\subsection{ Bipolaritons in GaAs-based microcavities }

The bipolariton model for excitonic molecules in $\lambda$-microcavities 
requires to construct the XX state in terms of quasi-bound 0$\lambda$-LB, 
1$\lambda$-LB, and 1$\lambda$-UB polaritons. In this case the radiative 
corrections to the XX state with ${\bf K}_{\|}\!=\!0$ are given by 
\begin{eqnarray}
\Delta^{\rm MC}_{\rm XX}({\bf K}_{\|}\!=\!0) &=& { 27 \over 8 } 
\sqrt{ \pi \over 2} \ \epsilon_{\rm XX}^{(0)} \ \mbox{Re} \Big\{ 
{ A \over 1 + B } \Big\} \, , 
\nonumber \\
\Gamma^{\rm MC}_{\rm XX}({\bf K}_{\|}\!=\!0) &=& - { 27 \over 4 } 
\sqrt{ \pi \over 2} \ \epsilon_{\rm XX}^{(0)} \ \mbox{Im} \Big\{ 
{ A \over 1 + B } \Big\} \, , 
\label{MCrad}
\end{eqnarray}
where
\begin{eqnarray}
A&=&{1 \over 2 \pi} \int_0^{+ \infty} \! \! \! \! \! \! \! p_{\|} dp_{\|}
\left[ {\tilde G}(p_{\|}) \left( \epsilon_{\rm XX}^{(0)} + {\hbar^2 
p_{\|}^2 \over M_x} \right) + 1 \right] \Psi_{\rm XX}^{(0)}(p_{\|}) \, , 
\nonumber \\
B&=&{ 27 \over 16 } {1 \over \sqrt{2 \pi} } \ \epsilon_{\rm XX}^{(0)} 
\int_0^{+ \infty} \! \! \! p_{\|} dp_{\|} \, {\tilde G}(p_{\|}) 
\Psi_{\rm XX}^{(0)}(p_{\|}) \, .
\label{mcAB}
\end{eqnarray}
In Eq.\,(\ref{mcAB}) the bipolariton Green function ${\tilde G}(p_{\|})$ 
is  
\begin{equation}
{\tilde G}(p_{\|}) =  \sum_{i,j} { [ u_i^{\rm MC}(p_{\|}) 
u_j^{\rm MC}(-p_{\|}) ]^2 \over E_{\rm XX}^{\rm MC} - 
\hbar \omega_i^{\rm MC}(p_{\|}) - \hbar \omega_j^{\rm MC}(-p_{\|}) 
+ i \gamma_0 } \, , 
\label{Green}
\end{equation}
where $E_{\rm XX}^{\rm MC} = 2E_{\rm X}(p_{\|}$=$0) - 
\epsilon_{\rm XX}^{(0)}$, the MC polariton eigenfrequency 
$\omega_{i(j)}^{\rm MC}$ and the X component $[u_{i(j)}^{\rm MC}]^2$ with 
$i,j =$ 0$\lambda$-LB, 1$\lambda$-LB, and 1$\lambda$-UB are given by 
Eq.\,(\ref{MC}) and Eq.\,(\ref{MCcomp}), respectively, and $\gamma_0 
\rightarrow +0$. The XX radiative corrections, i.e., the Lamb shift 
$\Delta_{\rm XX}^{\rm MC}$ and the radiative width 
$\Gamma_{\rm XX}^{\rm MC}$, depend upon the relative motion of the 
constituent QW excitons over whole momentum space, i.e., 
Eqs.\,(\ref{MCrad})-(\ref{mcAB}) include integration over $d p_{\|}$. 
The change of the input XX binding energy, $\epsilon_{\rm XX}^{(0)} 
\rightarrow \epsilon^{\rm XX} = \epsilon_{\rm XX}^{(0)} - 
\Delta_{\rm XX}^{\rm MC}({\bf K}_{\|}) + (i/2)   
\Gamma_{\rm XX}^{\rm MC}({\bf K}_{\|})$, occurs because in their relative 
motion the constituent excitons move along the MC polariton dispersion 
curves, rather than possess the quadratic dispersion, $E_{\rm X} = \hbar 
\omega_t + \hbar^2 p_{\|}^2/(2M_x)$ (the latter is valid only for 
optically inactive excitons). 

The solution of the exactly-solvable bipolariton model, given by 
Eqs.\,(\ref{MCrad})-(\ref{Green}), includes all possible channels of the 
in-plane dissociation of the microcavity molecule into two outgoing MC 
polaritons, i.e., ``XX (${\bf K}_{\|}$=0) $\rightarrow$ 
$i$th-branch MC polariton ($\sigma^+, {\bf p}_{\|}$) $+$ $j$th-branch MC 
polariton ($\sigma^-, -{\bf p}_{\|}$)''. Note that the solution of the 
bipolariton wave Eq.\,(\ref{BPint}) for excitonic molecules in a single QW 
can be obtained from Eqs.\,(\ref{MCrad})-(\ref{Green}) by putting $i = j$ 
= IP and replacing $u^{\rm MC}_{i(j)}$ and $\omega_{i(j)}^{\rm MC}$ by 
$u_{\rm IP}$ and $\omega_{\rm IP}$, respectively. 

\begin{figure}
\includegraphics*[width=8cm]{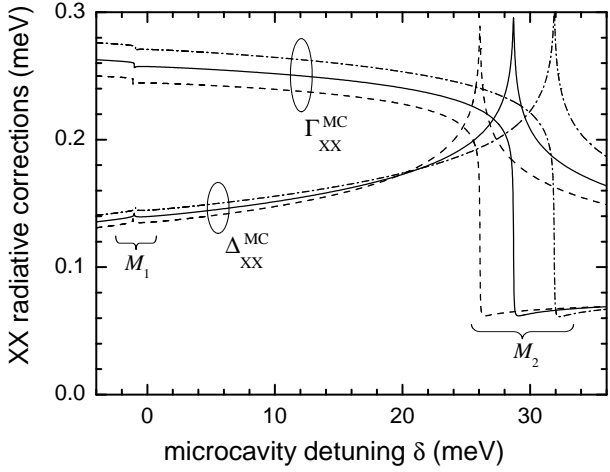}
\caption[]{ The radiative corrections to the excitonic molecule state, 
$\Gamma^{\rm MC}_{\rm XX}$ and $\Delta^{\rm MC}_{\rm XX}$, calculated 
against the MC detuning $\delta$ with Eqs.\,(\ref{MCrad})-(\ref{Green}) 
for the MC Rabi energies $\hbar \Omega^{\rm MC}_{1\lambda} = 7.76$\,meV 
and $\hbar \Omega^{\rm MC}_{0\lambda} = 5.60$\,meV. The input XX 
binding energy $\epsilon^{(0)}_{\rm XX}$ is 0.9\,meV (dash-dotted line), 
1.0\,meV (solid line), and 1.1\,meV (dashed line). }
\end{figure}

The radiative width $\Gamma_{\rm XX}^{\rm MC (1)} = 
\Gamma_{\rm XX}^{\rm MC}(\delta)$ and the Lamb shift 
$\Delta^{\rm MC}_{\rm XX} = \Delta^{\rm MC}_{\rm XX}(\delta)$ calculated 
by Eqs.\,(\ref{MCrad})-(\ref{Green}) as a function of the MC detuning 
$\delta = \hbar (\omega_0 - \omega_t)$ between the 1$\lambda$ cavity mode 
and QW exciton are plotted in Fig.\,4 for three values of the input XX 
binding energy, $\epsilon_{\rm XX}^{(0)} = 0.9$\,meV, 1.0\,meV, and 
1.1\,meV. By applying Eq.\,(\ref{strR}), we estimate for this plot the 
Rabi frequencies, $\Omega^{\rm MC}_{1\lambda}$ and 
$\Omega^{\rm MC}_{0\lambda}$, relevant to the used three-branch MC 
polariton dispersion given by Eq.\,(\ref{MC}). Namely, for $\hbar^2 
R^{\rm QW}_{\rm X} = 0.035\,\mbox{eV}^2 \AA$, associated with the 
reference QW, and $L_z = 2326\,\AA$, Eq.\,(\ref{strR}) yields 
$\hbar \Omega^{\rm MC}_{1\lambda} \simeq 7.76$\,meV and $\hbar 
\Omega^{\rm MC}_{0\lambda} \simeq 5.60$\,meV. As a result of non-ideal 
optical confinement in the $z$-direction by DBRs, our GaAs-based 
$\lambda$-microcavity (i) has a smaller value of 
$\Omega^{\rm MC}_{1\lambda}$, i.e., $\hbar \Omega^{\rm MC}_{1\lambda} 
\simeq 3.7$\,meV and (ii) with increasing $p_{\|}$ loses the strength 
of optical confinement for MC 1$\lambda$-mode photons of frequency 
$\omega^{\gamma} \simeq \omega_0$. The latter means that the MC photon 
radiative width $\gamma_{\rm R}$ is $p_{\|}$-dependent and smoothly 
increases with increasing $p_{\|}$. The DBR optical confinement is 
completely relaxed for $p_{\|} \sim p_0$ so that the dispersion 
Eq.\,(\ref{MC}) becomes inadequate, and the microcavity 0$\lambda$-LB 
polariton dispersion evolves towards the interface polariton dispersion, 
associated with the single QW and given by Eq.\,(\ref{pol}). Thus, in 
order to model the experimental data with 
Eqs.\,(\ref{MCrad})-(\ref{Green}), we use $\hbar 
\Omega^{\rm MC}_{1\lambda} \simeq 3.70$\,meV and $\hbar 
\Omega^{\rm MC}_{0\lambda} \simeq 2.67$\,meV, and replace the 
0$\lambda$-LB polariton dispersion by the interface, QW polariton 
dispersion with $\hbar^2 R^{\rm QW}_{\rm X} = 0.035\,\mbox{eV}^2 \AA$. 
For this case the plot of $\Gamma_{\rm XX}^{\rm MC}$ and 
$\Delta^{\rm MC}_{\rm XX}$ against the detuning $\delta$ is shown in 
Fig.\,10 (for details see Section IV). 

There are two sharp spikes in the dependence $\Delta^{\rm MC}_{\rm XX} = 
\Delta^{\rm MC}_{\rm XX}(\delta)$ which are accompanied by the 
jump-like changes of the XX radiative width $\Gamma^{\rm MC}_{\rm XX} = 
\Gamma^{\rm MC}_{\rm XX}(\delta)$ (see Figs.\,4 and 10). The above 
structure is due to van Hove critical points, $M_1$ and $M_2$, in the 
joint density of the polariton states (JDPS) relevant to the optical 
decay ``MC excitonic molecule ${\bf K}_{\|}$=0 $\rightarrow$ MC polariton 
${\bf p}_{\|}$ + MC polariton $-{\bf p}_{\|}$'' (for the critical points 
we use the classification and notations proposed in 
Ref.\,[\onlinecite{vanHove}]). The first critical point $M_1$ in 
energy-momentum space $\{ \delta, {\bf p}_{\|} \}$ refers to a negative 
MC detuning $\delta_1$ and deals with the condition $\mbox{Re} 
\{{\tilde E}^{\rm MC}_{\rm XX}({\bf K}_{\|}$=0$)\} = 2 \hbar \omega_t - 
\epsilon_{\rm XX}^{(0)} + \Delta^{\rm MC}_{\rm XX} = \hbar 
\omega^{\rm MC}_{\rm 1\lambda LB}({\bf p}_{\|}$=0$) + \hbar 
\omega^{\rm MC}_{\rm 1\lambda UB}(-{\bf p}_{\|}$=0). This point is 
marginal for the optical decay ``XX $\rightarrow$ 1$\lambda$-LB 
polariton + 1$\lambda$-UB polariton'': For $\delta \leq \delta_1$ 
the above channel is allowed, while it is absent for $\delta > 
\delta_1$. The critical point $M_2$ occurs at a positive detuning 
$\delta_2$, which corresponds to the condition $\mbox{Re} 
\{{\tilde E}^{\rm MC}_{\rm XX}({\bf K}_{\|}$=0$)\} = 2 \hbar \omega_t - 
\epsilon_{\rm XX}^{(0)} + \Delta^{\rm MC}_{\rm XX} = \hbar 
\omega^{\rm MC}_{\rm 1\lambda LB}({\bf p}_{\|}$=0$) + \hbar 
\omega^{\rm MC}_{\rm 1\lambda LB}(-{\bf p}_{\|}$=0), and is the main 
marginal point in the JDPS for the XX optical dissociation into two 
outgoing 1$\lambda$-LB polaritons. Namely, for $\delta \leq \delta_2$ 
the molecule can decay into two 1$\lambda$-LB polaritons, while for 
$\delta > \delta_2$ the optical decay of MC molecules with zero in-plane 
wavevector ${\bf K}_{\|}$ into $1\lambda$-LB polaritons is completely 
forbidden. With a very high accuracy of the order of $|\delta|/\omega_t 
\ll 1$, one finds from Eq.\,(\ref{MC}) that $\hbar \omega_{\rm 1\lambda 
UB/1\lambda LB}(p_{\|}$=$0) = \hbar \omega_t + (\delta/2) \pm (1/2) [ 
\delta^2 + (\hbar \Omega_{1 \lambda}^{\rm MC})^2 ]^{1/2}$. Thus from the 
energy-momentum conservation law we estimate the detunings $\delta_{1,2}$: 
\begin{eqnarray}
\! \! \! \! \! \! \! \!
\mbox{Critical point} \ M_1:&& \delta_1 = - \epsilon_{\rm XX}^{\rm MC} \, , 
\nonumber \\
\! \! \! \! \! \! \! \!
\mbox{Critical point} \ M_2:&& \delta_2 = { (\hbar 
\Omega_{1 \lambda}^{\rm MC})^2 - (\epsilon_{\rm XX}^{\rm MC})^2 \over 
2 \epsilon_{\rm XX}^{\rm MC} } \, , 
\label{vanHove}
\end{eqnarray}
where $\epsilon_{\rm XX}^{\rm MC} = \epsilon_{\rm XX}^{(0)} - 
\Delta_{\rm XX}^{\rm MC}$ is the true, ``measured'' binding energy of the 
bipolariton state ${\bf K}_{\|}$=0, i.e., of the optically dressed 
molecule. 

\begin{figure}
\includegraphics*[width=8cm]{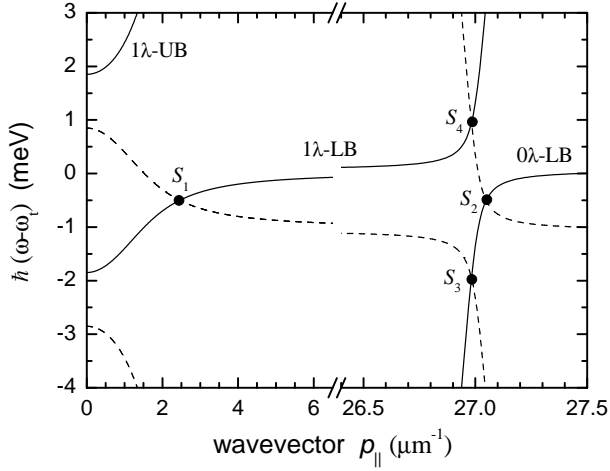}
\caption[]{ The graphic solution of the energy-momentum conservation law 
for the optical decay of a MC molecule with ${\bf K}_{\|}=0$. The 
microcavity Rabi energies are $\hbar \Omega^{\rm MC}_{1\lambda} = 
3.70$\,meV and $\hbar \Omega^{\rm MC}_{0\lambda} = 2.67$\,meV, the MC 
detuning is zero. The solutions are shown by the bold points $S_1$ 
(XX $\rightarrow$ 1$\lambda$-LB polariton + 1$\lambda$-LB polariton), 
$S_2$ (XX $\rightarrow$ 0$\lambda$-LB polariton + 0$\lambda$-LB 
polariton), and $S_{3,4}$ (XX $\rightarrow$ 0$\lambda$-LB polariton + 
1$\lambda$-LB polariton). The efficiency of the last decay channel 
is negligible in comparison with that of the first two. 
The XX binding energy $\epsilon^{(0)}_{\rm XX} = 1$\,meV. } 
\end{figure}

\begin{figure}
\includegraphics*[width=8cm]{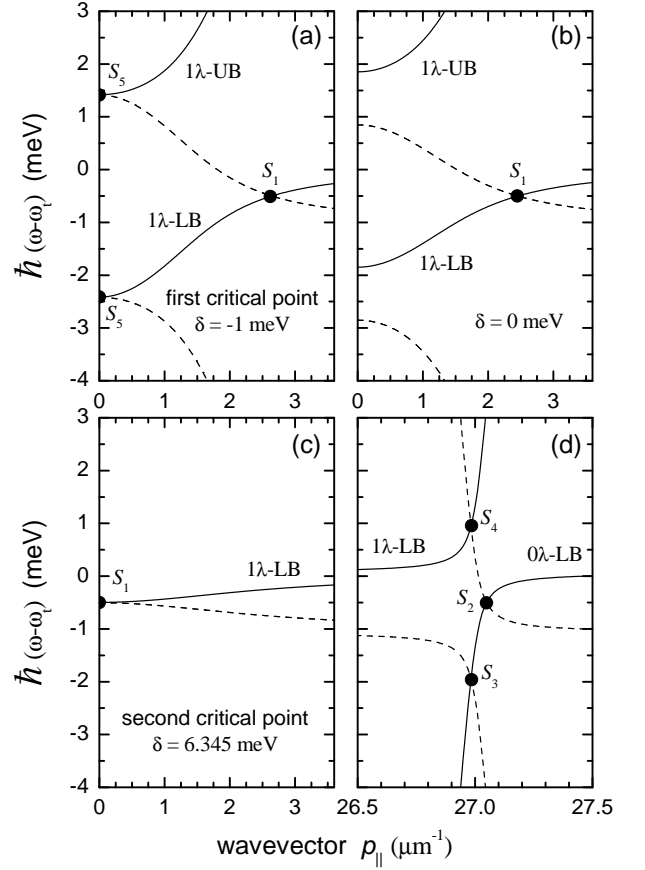}
\caption[]{ The graphic solution of energy-momentum conservation for 
(a) the decay channels ``XX $\rightarrow$ 1$\lambda$-LB polariton + 
1$\lambda$-LB polariton'' and ``XX $\rightarrow$ 1$\lambda$-LB 
polariton + 1$\lambda$-UB polariton'' (the MC detuning $\delta=-1$\,meV, 
the marginal solution $S_5$ at $p_{\|}=0$ refers to the critical point 
$M_1$); (b) the decay path ``XX $\rightarrow$ 1$\lambda$-LB polariton + 
1$\lambda$-LB polariton'' (the MC detuning $\delta=0$); (c) the decay 
path ``XX $\rightarrow$ 1$\lambda$-LB polariton + 1$\lambda$-LB 
polariton'' (the MC detuning $\delta=6.345$\,meV, the marginal solution 
$S_1$ at $p_{\|}=0$ refers to the critical point $M_2$); (d) the decay 
channels ``XX $\rightarrow$ 0$\lambda$-LB polariton + 0$\lambda$-LB 
polariton'' and ``XX $\rightarrow$ 0$\lambda$-LB polariton + 
1$\lambda$-UB polariton'' (this plot is practically independent of 
$\delta$). The MC Rabi frequencies $\Omega^{\rm MC}_{1\lambda}$ and 
$\Omega^{\rm MC}_{0\lambda}$, and the XX binding energy 
$\epsilon^{(0)}_{\rm XX}$ are the same as in Fig.\,5. } 
\end{figure}

In order to visualize the optical decay channels of MC excitonic 
molecules, in Figs.\,5 and 6 we plot the graphic solution of the 
energy-momentum conservation law, $E^{\rm MC}_{\rm XX} - \hbar 
\omega_i^{\rm MC}(p_{\|}) - \hbar \omega_j^{\rm MC}(-p_{\|}) = 0$ 
($i,j =$ 0$\lambda$-LB, 1$\lambda$-LB, and 1$\lambda$-UB). The 
roots of the equation are the poles of the bipolariton Green 
function ${\tilde G}$ given by Eq.\,(\ref{Green}). Figure 5, which 
refers to the zero-detuning GaAs-based microcavity, clearly 
illustrates that apart from the decay path ``XX $\rightarrow$ 
1$\lambda$-LB polariton + 1$\lambda$-LB polariton'' there are also the 
decay routes which involve the 1$\lambda$-LB and 0$\lambda$-LB 
microcavity polaritons with $p_{\|} \sim p_0$, i.e., ``XX $\rightarrow$ 
0$\lambda$-LB polariton + 0$\lambda$-LB polariton'' and 
``XX $\rightarrow$ 1$\lambda$-LB polariton + 0$\lambda$-LB polariton''. 
The graphic solution of energy-momentum conservation for the wavevector 
domain $p_{\|} \lesssim p_{\|}^{(1 \lambda)}$ is shown in a magnified 
scale in Figs.\,6a-6c for $\delta = \delta_1$, 0, and $\delta_2$, 
respectively. The touching points at $p_{\|}=0$ between the 
1$\lambda$-upper and 1$\lambda$-lower (see Fig.\,6a) and 
1$\lambda$-lower and 1$\lambda$-lower  (see Fig.\,6c) dispersion 
branches correspond to the $M_1$ and $M_2$ critical points, respectively. 
The graphic solution of the energy-momentum conservation law is shown in 
Fig.\,6d for the vicinity of $p_{\|} = p_0$. According to Eq.\,(\ref{MC}), 
the 1$\lambda$-LB and 0$\lambda$-LB polaritons with $p_{\|} \sim p_0$ 
practically do not depend upon the MC detuning $\delta$, i.e., the plot 
shown in Fig.\,6d is not sensitive to $\delta$. 

The value of the $\Gamma^{\rm MC}_{\rm XX}$-jump and 
$\Delta^{\rm MC}_{\rm XX}$-spike nearby the critical point $M_1$, i.e., 
at $\delta = \delta_1$, shows that the contribution of the decay path 
``XX $\rightarrow$ 1$\lambda$-UB polariton + 1$\lambda$-LB polariton'' 
is rather small, about 1-2$\%$ only. This is mainly due to a small 
value of the JDPS in the decay channel. The main contribution to the 
XX radiative corrections in microcavities is due to the 
frequency-degenerate decay routes ``XX $\rightarrow$ 1$\lambda$-LB 
polariton + 1$\lambda$-LB polariton'' and ``XX $\rightarrow$ 
0$\lambda$-LB polariton + 0$\lambda$-LB polariton'' (or ``XX 
$\rightarrow$ interface polariton + interface polariton'', as a result 
of the relaxation of the transverse optical confinement at $p_{\|} \sim 
p_0$). The JDPS associated with the first main channel is given by 
\begin{eqnarray}
\rho^{\rm XX \rightarrow 1\lambda LB + 1\lambda LB}_{\rm \hbar \omega 
= \hbar \omega_t - \epsilon^{\rm MC}_{\rm XX}/2} &=& {\pi \over 2 \hbar 
\omega_t} \left( a_{\rm XX}^{\rm (2D)} p_0 \right )^2 
\nonumber \\ 
\times \Bigg[ 1 &+& \left( { \Omega^{\rm MC}_{\rm 1 \lambda } 
\over \epsilon_{\rm XX}^{(0)} } \right)^2 \Bigg] \Theta( \delta_2 - 
\delta ) \, , 
\label{JDPS}
\end{eqnarray}
where $\Theta(x)$ is the Heaviside step function. The above JDPS is 
relevant to the calculations done by the bipolariton 
Eqs.\,(\ref{MCrad})-(\ref{Green}). The appearance of the dimensionless 
parameter $\delta_{\rm R}^{\rm (2D)} = ( a^{\rm (2D)}_{\rm XX} p_0 )^2$ 
on the right-hand side (r.h.s.) of Eq.\,(\ref{JDPS}) is remarkable. Thus 
the same control parameter $\delta_{\rm R}^{\rm (2D)}$ determines the 
optical decay of excitonic molecules in the reference single GaAs QW and 
in the GaAs-based microcavities. Furthermore, the JDPS given by 
Eq.\,(\ref{JDPS}) depends upon the MC detuning only through the step 
function $\Theta ( \delta_2 - \delta )$. The latter dependence gives 
rise to the critical point $M_2$. By comparing the XX radiative 
corrections for $\delta < \delta_2$ and $\delta > \delta_2$ (see 
Figs.\,4 and 10), one concludes that the first main decay channel ``XX 
$\rightarrow$ 1$\lambda$-LB polariton + 1$\lambda$-LB polariton'' has 
nearly the same efficiency as the second one, ``XX $\rightarrow$ 
0$\lambda$-LB polariton + 0$\lambda$-LB polariton'' (or ``XX 
$\rightarrow$ interface polariton + interface polariton''). Note that 
the ``virtual'' decay paths, like ``XX $\rightarrow$ 1$\lambda$-UB 
polariton + 1$\lambda$-UB polariton'', still contribute to the XX Lamb 
shift in microcavities, according to Eqs.\,(\ref{MCrad})-(\ref{Green}).

\section{Experiment}

The investigated sample consists of an MBE--grown 
GaAs/Al$_{0.3}$Ga$_{0.7}$As single quantum well of the thickness $d_z = 
250\,\AA$ and placed in the center of a $\lambda$--cavity. An 
AlAs/Al$_{0.15}$Ga$_{0.85}$As DBR of 25 (16) periods was grown at the 
bottom (top) of the cavity. The spacer layer is wedged, in order to tune 
the cavity mode along the position on the sample. Details on the growth 
and sample design can be found in Ref.\,[\onlinecite{Jacob00}]. The 
optical properties of the reference single QW grown under nominally 
identical conditions are reported in Ref.\,[\onlinecite{LangbeinPRB2000}]: 
The spectra show the ground-state heavy-hole (HH) and light-hole (LH) 
exciton absorption lines separated in energy by about 2.6\,meV. In the 
MC sample, the coupling of both HH and LH excitons with the 
1$\lambda$-mode cavity photons results in the formation of three 
1$\lambda$-eigenmode MC polariton dispersion branches, 1$\lambda$-LB, 
1$\lambda$-MB, and 1$\lambda$-UB \cite{Jacob00}. The 1$\lambda$-mode 
polaritons have a narrow linewidth: The ratio between the HH Rabi 
splitting and the polariton linewidths at zero detuning is about twenty 
\cite{BorriPRB2000}. 

For the reference GaAs QW at temperature $T \lesssim 10$\,K the 
homogeneous width ${\tilde \Gamma}^{\rm QW}_{\rm X}$ is dominated by 
the radiative decay. The absorption linewidth, measured along the 
$z$-direction and extrapolated to zero temperature, yields the HH--X 
radiative width of $98 \pm 10\,\mu$eV. Note that this value is affected 
by optical interference which occurs at the position of the QW, $z=0$, 
due to bulk photons emitted by the QW excitons and partly reflected back 
by the top surface ($z = L_{\rm cap} \simeq 499$\,nm) of a cap layer. In 
this case one has a constructive interference which results in the 
enhancement of the light field at $z=0$. By treating the optical 
interference effect, we estimate $\Gamma_{\rm X}^{\rm QW} \simeq 60\,\mu 
{\rm eV}$ for the reference QW sandwiched between semi-infinite bulk 
AlGaAs barriers. This radiative width yields the intrinsic oscillator 
strength of quasi-2D HH excitons $\hbar^2 
R_{\rm X}^{\rm QW}(d_z$=$250\,\AA) \simeq 0.035\,{\rm eV}^2\AA$ (see 
Fig.\,3). The measured characteristics of excitonic molecules in the 
reference QW are consistent with those reported in 
Ref.\,[\onlinecite{LangbeinPRB2000}]: The XX binding energy 
$\epsilon^{\rm MC}_{\rm XX} \simeq 0.9-1.1$\,meV and the XX radiative 
width $\Gamma^{\rm MC}_{\rm XX} \simeq 0.1$\,meV. The latter value is 
obtained by extrapolating the measured homogeneous width 
${\tilde \Gamma}^{\rm MC}_{\rm XX} = 
{\tilde \Gamma}^{\rm MC}_{\rm XX}(T)$ to $T = 0$\,K. 

The optical experiments with the MC sample were performed using a 
Ti:sapphire laser source which generates Fourier-limited 100\,fs laser 
pulses at 76\,MHz repetition rate. Two exciting pulses, 1 and 2, with 
variable relative delay time $\tau_{12}$ propagate along two different 
incident directions ${\bf p}_{1,2}$ at small angle ($\leq1^{\circ}$) to 
the surface normal. Pulse 1 precedes pulse 2 for $\tau_{12}>0$. The  
reflectivity spectra of the probe light and the FWM signal were analyzed 
with a spectrometer and a charge-coupled device camera of 140\,$\mu$eV 
FWHM resolution. The sample was held in a Helium bath cryostat at T=5\,K 
for all the pump-probe measurements and at T=9\,K in the FWM experiments.

\subsection{Bipolariton dephasing in GaAs microcavities}

\begin{figure}
\includegraphics*[width=8cm]{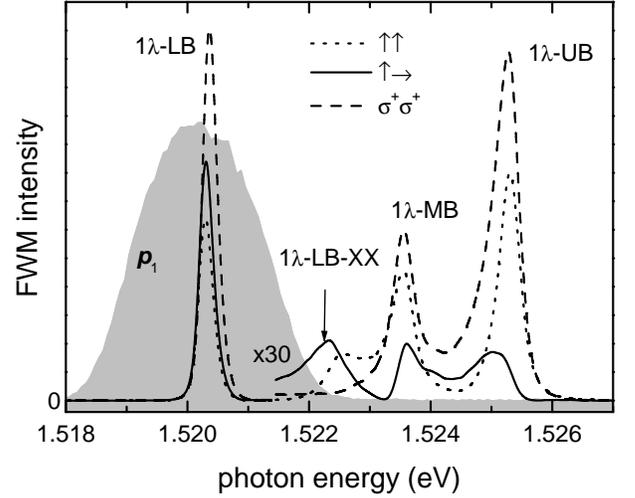}
\caption[]{ Spectrally resolved four--wave mixing for co-circular 
(dashed line), co-linear (bold dotted line), and cross-linear (solid 
line) polarizations of the exciting pulses. The microcavity detuning 
is $\delta = 0.76$\,meV. The pulse along ${\bf p}_1$-direction induces 
only 1$\lambda$-LB polaritons, and its spectrum is shown by the dotted 
line. } 
\end{figure}

In order to measure the bipolariton dephasing we perform spectrally 
resolved FWM. The FWM signal was detected at $2 {\bf p}_2 - {\bf p}_1$ 
in reflection geometry. The spot size of both exciting beams was $\sim 
50$\,$\mu$m. In Fig.\,7 we plot the spectrally--resolved FWM signal for 
different polarization configurations of the laser pulses. The positive 
detuning between the cavity 1$\lambda$-eigenmode and HH exciton is 
$\delta = 0.76$\,meV, and the delay time is $\tau_{12}$=1\,ps. Pulse 1 
of about 500\,fs duration was spectrally shaped to {\it excite only the 
1}$\lambda$-{\it LB polaritons}, and the FWM was probed with the 
spectrally broad pulse 2 at all 1$\lambda$-mode polariton resonances. 
For co--linear and cross-linear polarization configurations, the 
1$\lambda$-LB polariton to excitonic molecule transition (1$\lambda$-LB 
-- XX) is observed in the FWM signal (see arrow in Fig.\,7) at a spectral 
position consistent with that found in our previous pump-probe 
experiments \cite{BorriPRBRb00}. The XX-mediated FWM signal disappears 
for co--circular polarization, in accordance with the polarization 
selection rules for the two-photon generation of excitonic molecules 
in a GaAs QW. 

Although the analysis of FWM in microcavities can be rather complicated 
\cite{Gonokami97,ShiranePRB98}, the interpretation of our measurements 
is simplified by the selective excitation of the 1$\lambda$-LB polaritons 
only. The observed TI--FWM is a free polarization decay, due to the 
dominant homogeneous broadening of the X lines in our high-quality 
250\,$\AA$-wide QWs \cite{LangbeinPRB2000}. At positive delays the FWM 
signal is created by the following sequence. At first, pulse 1 induces 
a first--order polarization associated with 1$\lambda$-LB polaritons.  
The induced polarization decays with the dephasing time 
$T_2^{\rm 1\lambda-LB}$ of the 1$\lambda$-LB polaritons. The dephasing 
time $T_2^{\rm 1\lambda-LB}$ is dominated by the lifetime of 
1$\lambda$-mode MC photons. Pulse 2 interacts nonlinearly with the 
induced polarization, and a third--order FWM signal is created with an 
amplitude that decreases with increasing $\tau_{12}$, due to the decay 
of the first--order polarization associated with the 1$\lambda$-LB 
polaritons. The TI--FWM intensities at all probed resonances therefore 
decay nearly with the time constant $T_{2}^{\rm 1\lambda-LB}$/2. At 
negative $\tau_{12}$ the FWM signal stems from the two--photon coherence 
of the crystal ground state to the excitonic molecule transition (0--XX) 
induced by pulse 2. According to energy -- in-plane momentum conservation, 
since pulse 1 is resonant with 1$\lambda$-LB polaritons only, the FWM 
signal, associated with bulk photons, is emitted in the direction 
$2{\bf p}_2 - {\bf p}_1$ with the energy of the 1$\lambda$-LB -- XX 
transition. Thus the TI--FWM dynamics at negative time delays allows us 
to study the polarization decay of the 0--XX transition \cite{BorriOECS02}, 
i.e., to find ${\tilde \Gamma}^{\rm MC}_{\rm XX}$. 

\begin{figure}
\includegraphics*[width=8cm]{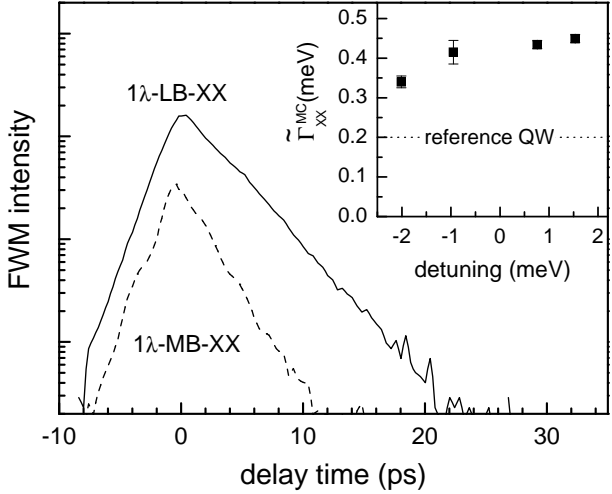}
\caption[]{ Comparison between the FWM dynamics measured at the 
1$\lambda$-LB -- XX transition, when pulse 1 resonantly induces the 
1$\lambda$-mode lower-branch polaritons only, and at the 1$\lambda$-MB -- 
XX transition, when pulse 1 resonantly excites only the 1$\lambda$-mode 
middle-branch polaritons. Inset: The XX homogeneous linewidth 
${\tilde \Gamma}^{\rm MC}_{\rm XX}$ against the MC detuning $\delta$, 
measured at $T=9$\,K with about 4\,nJ/cm$^{2}$ pump fluence. }
\end{figure}

The $\tau_{12}$-dependence of the TI--FWM signals associated with the 
1$\lambda$-LB -- XX and 1$\lambda$-MB -- XX transitions is shown in 
Fig.\,8. As expected, at negative $\tau_{12}$ one finds the same 
dynamics for both transitions. Therefore, independently of the 
1$\lambda$-eigenmode MC polariton branch selectively excited by pulse 1, 
we can infer the polarization decay rate of the 0--XX transition. The 
homogeneous linewidth of the 0--XX transition 
${\tilde \Gamma}^{\rm MC}_{\rm XX}$ measured at low excitation energies 
per pulse ($\sim4$\,nJ/cm$^{2}$) is potted against the MC detuning 
$\delta$ in the inset of Fig.\,8. Only a weak detuning dependence of 
${\tilde \Gamma}^{\rm MC}_{\rm XX}$ is observed for the detuning band 
$-2\,\mbox{meV} \leq \delta \leq 2\,\mbox{meV}$. Note that the deduced 
values ${\tilde \Gamma}^{\rm MC}_{\rm XX}(T$=9\,K) $\simeq 0.3-0.4$\,meV 
are by factor $1.5-2$ larger than ${\tilde \Gamma}^{\rm QW}_{\rm XX} 
\simeq 0.2$\,meV measured from the reference QW at nearly the same bath 
temperature T=10\,K \cite{LangbeinPRB2000} (see the dotted line in the 
inset of Fig.\,8).

\subsection{The binding energy of bipolaritons in GaAs microcavities}

The bipolariton energy $E_{\rm XX}^{\rm MC}$ was found by analysing the 
pump-probe experiments. Pulse 1 acts as an intense pump while pulse 2 
is a weak probe. The spectrum of the pump pulse is shaped and tuned in 
order to excite resonantly the 1$\lambda$-LB polaritons only. The 
spectrally broad probe pulse has a spot size of $\sim 40$\,$\mu$m. In 
this case the in-plane spatial gradient of the polariton energy is not 
significant. In order to achieve a uniform pump density over the probe 
area, the cross-section of the pump pulse is chosen to be by factor 
two larger than that of the probe light. 

In Ref.\,[\onlinecite{BorriPRBRb00}] we show a well-resolved 
pump-induced absorption at the 1$\lambda$-LB -- XX transition 
in the investigated MC sample. The 1$\lambda$-LB -- XX absorption 
was observed in the reflectivity spectra at positive pump-probe delay 
times and for the cross-circularly ($\sigma^+$- and $\sigma^-$-) 
polarized pump and probe pulses, according to the optical selection 
rules. In particularly, the induced absorption for three different 
positive MC detunings was measured. Here we extend the 
pump-probe experiment to study the detuning dependence 
$E_{\rm XX}^{\rm MC} = E_{\rm XX}^{\rm MC}(\delta)$, including $\delta 
< 0$. In Fig.\,9 the probe reflectivity spectra measured at $\tau_{12} 
\simeq 0.5$\,ps for the cross-circularly polarized pump and probe 
pulses is plotted. Indicated by the arrows (see Fig.\,9), a spectrally 
well-resolved pump-induced absorption resonance is observed. In the 
upper left-hand side (l.h.s.) part of Fig.\,9 the energy position of 
the 1$\lambda$-LB, 1$\lambda$-MB, and 1$\lambda$-UB 
polariton resonances and of the induced 1$\lambda$-LB -- XX absorption 
are plotted against the MC detuning $\delta$. The fit done with a  
three-coupled-oscillator scheme (1$\lambda$-eigenmode MC photon, 
HH exciton, and LH exciton resonances) are shown by the solid lines. The 
energies $E^{\rm HH}_{\rm X}$ and $E^{\rm LH}_{\rm X}$ of the HH and 
LH excitons ($E^{\rm HH}_{\rm X} \simeq 1.5219$\,eV and 
$E^{\rm LH}_{\rm X} \simeq 1.5245$\,eV) are inferred from the fit, and 
the molecule energy $E^{\rm MC}_{\rm XX}$ is determined as the sum of 
the measured 1$\lambda$-LB and 1$\lambda$-LB -- XX transition energies. 
The bipolariton binding energy, evaluated as $\epsilon^{\rm MC}_{\rm XX} 
= 2 E^{\rm HH}_{\rm X} - E^{\rm MC}_{\rm XX}$, is plotted against the MC 
detuning $\delta$ in the lower l.h.s. part of Fig.\,9. We find that 
$\epsilon_{\rm XX}^{\rm MC} \simeq 0.9-1.1$\,meV, i.e., is similar to 
the value of $\epsilon_{\rm XX}^{\rm QW}$ in the reference single QW 
and slightly larger than that previously reported in 
Ref.\,[\onlinecite{BorriPRBRb00}]. 

\begin{figure}
\includegraphics*[width=8cm]{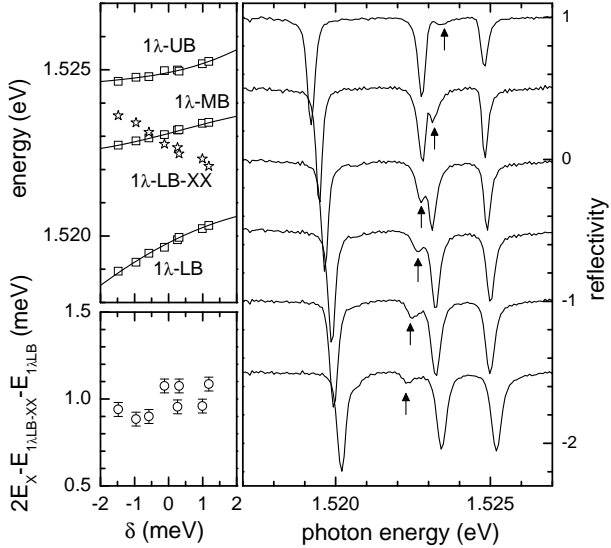}
\caption[]{ The reflectivity spectra of the probe light at different 
detuning values. The spectra are measured for the cross-circularly 
polarized pump and probe pulses at delay time $\tau_{12} \simeq 0.5$\,ps 
and $T=5$\,K. The pump fluence is about 0.1\,$\mu$J/cm$^{2}$. The arrows 
indicate the 1$\lambda$-LP -- XX pump-induced absorption. Upper left 
inset: The measured energy position of the 1$\lambda$-LP, 1$\lambda$-MP, 
1$\lambda$-UP resonances (filled square points) and of the induced 
1$\lambda$-LP -- XX absorption (unfilled square points) versus the MC 
detuning $\delta$. The fit of the MC polariton branches, associated 
with LH and HH QW excitons, is shown by the solid lines. Lower left 
inset: The XX binding energy $\epsilon^{\rm MC}_{\rm XX}$ determined 
as the difference between twice the bare HH exciton energy and the 
sum of the 1$\lambda$-LP and 1$\lambda$-LP -- XX transition energies. }
\end{figure}

\section{Discussion}

The optical decay of MC bipolaritons can also occur directly, through 
escape of the photon component of the constituent $\sigma^+$- and 
$\sigma^-$-polarized MC polaritons into the bulk photon modes. The XX 
radiative width associated with this channel is given by 
\begin{eqnarray}
\Gamma_{\rm XX}^{\rm MC (2)} ({\bf K}_{\|}\!\!&=&\!\!0) = {\hbar \over 
\pi} \gamma_{\rm R}  \int_0^{\infty} |\Psi_{\rm XX}^{(0)}(2p_{\|})|^2 
\nonumber \\
&\times& \Big[ \sum_i \left[ 1 - (u_i^{\rm MC})^2 \right] \Big] p_{\|} 
dp_{\|} \, ,
\label{MCbulk}
\end{eqnarray}
where $u_i^{\rm MC} = u_i^{\rm MC}(p_{\|})$ are determined by 
Eq.\,(\ref{MCcomp}) and $i$ runs over 0$\lambda$-LB, 1$\lambda$-LB, and 
1$\lambda$-UB. Equation (\ref{MCbulk}) is akin to Eq.\,(\ref{rada}) and 
can be interpreted in terms of optical evaporation of the MC excitonic 
molecules through the DBR mirrors. Using the measured radiative 
linewidth of 1$\lambda$-LB polaritons, 
$\Gamma_{\rm X}^{\rm MC}(p_{\|}\!\simeq\!0.13\!\times\!10^5\,{\rm cm}^{-1}) 
\simeq 0.1$\,meV, we estimate $\hbar \gamma_{\rm R} \simeq 0.3$\,meV, so 
that the radiative lifetime of MC photons is given by $\tau_{\rm R} 
\simeq 2.4$\,ps. In this case Eq.\,(\ref{MCbulk}) yields 
$\Gamma_{\rm XX}^{\rm MC (2)} \simeq 1 - 2\,\mu$eV for 
$\epsilon_{\rm XX}^{(0)} \simeq 0.9 - 1.1$\,meV and assuming that 
$\gamma_{\rm R}$ is $p_{\|}$-independent. Thus 
$\Gamma_{\rm XX}^{\rm MC (2)}$ is less than 
$\Gamma_{\rm XX}^{\rm QW (2)}$, estimated with Eq.\,(\ref{rada}) for 
the reference QW (see Fig.\,3), by more than one order of magnitude. 
This is because instead of the smallness parameter 
$\delta^{\rm (2D)}_{\rm R} = (a^{\rm (2D)}_{\rm XX} p_0)^2$, which 
appears on the r.h.s. of Eq.\,(\ref{rada}), Eq.\,(\ref{MCbulk}) is scaled 
by $(a^{\rm (2D)}_{\rm XX} p_{\|}^{(1\lambda)})^2 \ll 
\delta^{\rm (2D)}_{\rm R}$. 

The radiative width $\Gamma_{\rm XX}^{\rm MC (2)}$, associated with 
the decay of XXs into the bulk photon modes, is by two orders of 
magnitude less than $\Gamma_{\rm XX}^{\rm MC (1)}$ calculated with 
Eqs.\,(\ref{MCrad})-(\ref{Green}). Thus the resonant in-plane 
dissociation of molecules into outgoing MC polaritons absolutely 
dominates in the XX-mediated optics of microcavities, so that the total 
XX radiative width is given by $\Gamma_{\rm XX}^{\rm MC} = 
\Gamma_{\rm XX}^{\rm MC (1)} + \Gamma_{\rm XX}^{\rm MC (2)} \simeq 
\Gamma_{\rm XX}^{\rm MC (1)}$ (see Figs.\,4 and 10). The extremely small 
value of $\Gamma_{\rm XX}^{\rm MC (2)}$ allows us to interpret a MC 
excitonic molecule as a nearly ``optically-dark'' state with respect 
to its direct decay into the bulk photon modes. However it is the 
resonant coupling between 1$\lambda$-mode cavity polaritons and external 
bulk photons which is responsible for the optical generation and probe 
of the XX states in microcavities: Our optical experiments deal only 
with bulk pump, probe, and signal photons. In the meantime the 
bipolariton wavefunction ${\tilde \Psi}_{\rm XX}$ is constructed in 
terms of 0$\lambda$-LB, 1$\lambda$-LB, and 1$\lambda$-UB polariton 
states, and umklapp between the MC polariton branches occurs through 
the coherent Coulombic scattering of two constituent polaritons. 

While the interpretation of the experimental data (see Section III) does 
require three-branch, 1$\lambda$-LB, 1$\lambda$-MB, and 1$\lambda$-UB, 
polaritons associated with HH and LH excitons, the contribution to the 
XX optics from the LH Xs is very small. This occurs because (i) the energy 
$E^{\rm LH}_{\rm X}$ is well-separated from the XX-mediated resonance at 
$E^{\rm HH}_{\rm X} - \epsilon^{\rm MC}_{\rm XX}/2$ (the relevant ratio 
between $\epsilon^{\rm MC}_{\rm XX}/2$ and $E^{\rm LH}_{\rm X} - 
E^{\rm HH}_{\rm X} + \epsilon^{\rm MC}_{\rm XX}/2$ is equal to 0.16, 
i.e., is much less than unity) and (ii) because a contribution of the LH 
exciton to the total XX wavefunction is unfavorable in energy, i.e., is 
rather minor. We have checked numerically that by the first argument only 
the LH-X resonance cannot change the XX radiative corrections for more 
than 3-5$\%$. Thus the bipolariton model we develop to analyze the 
optical properties of MC excitonic molecules and to explain the 
experimental data deals only with 0$\lambda$-LB, 1$\lambda$-LB, and 
1$\lambda$-UB polaritons associated with the ground-state HH exciton. 

\begin{figure}
\includegraphics*[width=8cm]{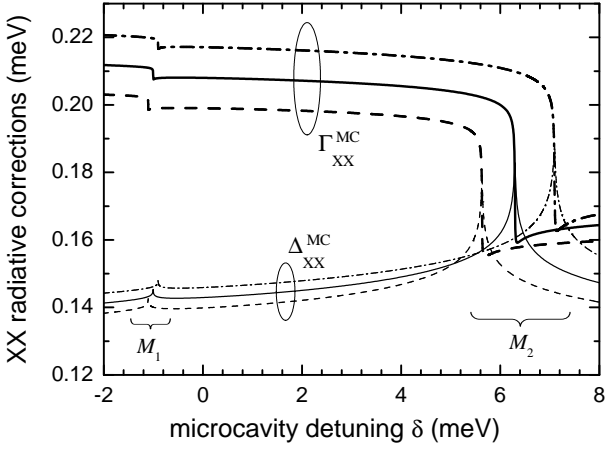}
\caption[]{ The radiative corrections to the excitonic molecule state, 
$\Gamma^{\rm MC}_{\rm XX}$ and $\Delta^{\rm MC}_{\rm XX}$, calculated 
versus the MC detuning $\delta$ with Eqs.\,(\ref{MCrad})-(\ref{Green}) 
for the MC Rabi energy $\hbar \Omega^{\rm MC}_{1\lambda} = 3.70$\,meV 
and assuming that the DBR optical confinement is completely relaxed for 
$p_{\|} \geq p_{\|}^{(1\lambda)} = 10^5\,\mbox{cm}^{-1}$. In this case  
the 0$\lambda$-LB dispersion is replaced by the interface polariton 
dispersion with the oscillator strength $\hbar^2 R^{\rm QW}_{\rm X} = 
0.035\,\mbox{eV}^2 \AA$. The input XX binding energy 
$\epsilon^{(0)}_{\rm XX}$ is 0.9\,meV (dash-dotted line), 1.0\,meV 
(solid line), and 1.1\,meV (dashed line). }
\end{figure}

In Fig.\,10 we plot the XX radiative corrections against the MC detuning 
$\delta$, calculated with Eqs.\,(\ref{MCrad})-(\ref{Green}) by using the 
MC parameters adapted to our GaAs microcavities. Namely, the 
1$\lambda$-mode cavity Rabi splitting is given by $\hbar 
\Omega^{\rm MC}_{1\lambda} = 3.7$\,meV, and we assume that the DBR 
optical confinement follows the step function $\Theta(p_{\|}^{(1\lambda)} 
- p_{\|})$. For $p_{\|} \geq p_{\|}^{(1\lambda)}$ the 
0$\lambda$-LB is replaced by the interface polariton dispersion given by 
Eq.\,(\ref{pol}). Due to the absence of the DBR transverse optical 
confinement at $p_{\|} \geq p_{\|}^{(1\lambda)}$, the resonant optical 
decay of the constituent excitons into the bulk photon mode is also 
included in our calculations by using Eq.\,(\ref{rada}) with integration 
over $dp_{\|}$ from $p_{\|}^{(1\lambda)}$ to $p_0$. From Fig.\,10 we 
conclude that for the detuning band $-2\,\mbox{meV} \lesssim \delta 
\lesssim 2\,\mbox{meV}$ the radiative width $\Gamma^{\rm MC}_{\rm XX}$ 
is about $0.20-0.22$\,meV and indeed weakly depends upon $\delta$, in 
accordance with our experimental data. A few $\mu$eV 
$\Gamma^{\rm MC}_{\rm XX}$-jump, associated with the critical point 
$M_1$, is too small to be detected in the current experiments. Note that 
the contribution to $\Gamma^{\rm MC}_{\rm XX}$ from the decay channel 
``XX $\rightarrow$ 1$\lambda$-LB polariton + 1$\lambda$-LB polariton'' 
can easily be estimated within a standard perturbation theory: 
$\Gamma^{\rm MC}_{\rm XX \rightarrow 1\lambda LB + 1\lambda LB} \simeq 
(\hbar \Omega^{\rm MC}_{1\lambda})^2 \rho^{\rm XX \rightarrow 1\lambda LB 
+ 1 \lambda LB}_{\hbar \omega = \hbar \omega_t - 
\epsilon^{\rm MC}_{\rm XX}/2}$, where the JDPS is given by 
Eq.\,(\ref{JDPS}). The above estimate yields $\Gamma^{\rm MC}_{\rm XX 
\rightarrow 1\lambda LB + 1\lambda LB} \simeq 0.06$\,meV and is 
consistent with the value of the $\Gamma^{\rm MC}_{\rm XX}$-jump around 
$\delta = \delta_2$, i.e., at the $M_2$ critical point (see Fig.\,10). An 
observation of $\Gamma^{\rm MC}_{\rm XX} \simeq 0.10-0.15$\,meV at 
$\delta > \delta_2$, when the MC excitonic molecules become 
optically dark with respect to the decay into 1$\lambda$-mode MC 
polaritons, would be a direct visualization of the hidden decay path 
``XX $\rightarrow$ interface polariton + interface polariton''. 

The {\it relative} change of the XX radiative corrections is rather small 
to be observed in the tested MC detuning band $|\delta| \leq 2$\,meV with 
the current accuracy of our measurements: 
Eqs.\,(\ref{MCrad})-(\ref{Green}) yield 
$\epsilon^{\rm MC}_{\rm XX}(\delta\!\!=\!\!2\,\mbox{meV}) - 
\epsilon^{\rm MC}_{\rm XX}(\delta\!\!=\!\!-2\,\mbox{meV}) \simeq - 
4\,\mu$eV and $\Gamma^{\rm MC}_{\rm XX}(\delta\!\!=\!\!2\,\mbox{meV}) - 
\Gamma^{\rm MC}_{\rm XX}(\delta\!\!=\!\!-2\,\mbox{meV}) \simeq - 
5\,\mu$eV; the energy structure at $\delta = \delta_1 = - 
\epsilon^{\rm MC}_{\rm XX}$, i.e., nearby the critical point $M_1$, is 
also of a few $\mu$eV only (see Fig.\,10). On the other hand, the 
GaAs-based microcavities we have now do not allow us to test the critical 
point $M_2$ which is located in the MC detuning band 5\,meV $< \delta <$ 
8\,meV. In the latter case the relative change of 
$\Delta_{\rm XX}^{\rm MC}$ and $\Gamma_{\rm XX}^{\rm MC}$ is large 
enough, about $0.04-0.07$\,meV, to be detected in our experiments. 
High-precision modulation spectroscopy is very relevant to observation 
of the critical points, because the derivatives $\partial^n 
(\Delta_{\rm XX}^{\rm MC}) / \partial \delta^n$ ($n \geq 1$) and 
$\partial^n (\Gamma_{\rm XX}^{\rm MC}) / \partial \delta^n$ ($n \geq 1$) 
undergo a sharp change in the spectral vicinity of $M_{1,2}$. The 
modulation of $\delta$ can be done by applying time-dependent 
quasi-static electric \cite{Fisher95}, magnetic \cite{Armitage97} or 
pressure \cite{Zhang02} fields. Note that the measurement of the 
detunings $\delta_1$ and $\delta_2$ will allow us to determine with a 
very high accuracy, by using Eqs.\,(\ref{vanHove}), the XX binding 
energy $\epsilon^{\rm MC}_{\rm XX}$ and the MC Rabi frequency 
$\Omega_{1\lambda}^{\rm MC}$. A detailed study of the XX Lamb shift 
$\Delta_{\rm XX}^{\rm MC}$ versus the MC detuning $\delta$ and, in 
particular, the detection of the critical points $M_1$ and $M_2$ are the 
issue of our next experiments. 

In order to estimate the radiative width $\Gamma^{\rm MC}_{\rm XX}$ from 
the total homogeneous width ${\tilde \Gamma}^{\rm MC}_{\rm XX}$ measured 
at $T = 9$\,K in our FWM experiment, we assume that apart from the XX 
radiative decay the main contribution to 
${\tilde \Gamma}^{\rm MC}_{\rm XX}$ is due to temperature-dependent XX 
-- LA-phonon scattering. Note that in the experiment we deal with a 
low-intensity limit, when ${\tilde \Gamma}^{\rm MC}_{\rm XX}$ is nearly 
independent of the excitation level. Thus 
${\tilde \Gamma}^{\rm MC}_{\rm XX} = \Gamma^{\rm MC}_{\rm XX} + 
\Gamma^{\rm QW}_{\rm XX-LA}$, where $\Gamma^{\rm QW}_{\rm XX-LA}$ is 
due to the scattering of QW molecules by bulk LA-phonons. The DBR optical 
confinement does not influence the XX -- LA phonon scattering, so that 
the width $\Gamma^{\rm QW}_{\rm XX-LA} = \Gamma^{\rm QW}_{\rm XX-LA}(T)$ 
is the same for XXs in the reference single QW and in the microcavities. 
$\Gamma^{\rm QW}_{\rm XX-LA}$ is given by 
\begin{eqnarray}
\Gamma^{\rm QW}_{\rm XX-LA}({\bf K}_{\|}\!=\!0) &=& 2 \pi {\hbar \over 
\tau_{\rm sc}} \int_1^{\infty} d \varepsilon \varepsilon 
\sqrt{ \varepsilon \over \varepsilon - 1 } 
\nonumber \\
&\times&
| F_z(a \sqrt{ \varepsilon (\varepsilon - 1)} |^2 
n^{\rm ph}_{\varepsilon} \, , 
\label{XX-LA}
\end{eqnarray}
where $\tau_{\rm sc} = (\pi^2 \hbar^4 \rho)/(32 D_x^2 M_x^3 v_s)$, $v_s$ 
is the longitudinal sound velocity, $D_x$ is the X deformation potential, 
$\rho$ is the crystal (GaAs) density, $n^{\rm ph}_{\varepsilon} = 
1/[\exp(\varepsilon E_0/k_{\rm B}T) - 1]$, and $E_0 = 4 M_x v_s^2$. The 
form-factor $F_z(x) = [\sin(x)/x][e^{ix}/(1 - x^2/\pi^2)]$ refers to an 
infinite rectangular QW confinement potential and describes the 
relaxation of the momentum conservation law in the $z$-direction. The 
dimensionless parameter $a$ is given by $a = (2 d_z M_x v_s)/\hbar$. 

\begin{figure}
\includegraphics*[width=8cm]{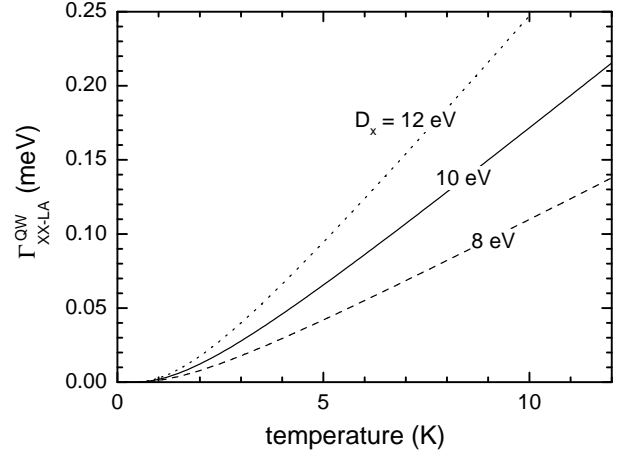}
\caption[]{ The temperature dependence of the homogeneous width 
$\Gamma^{\rm QW}_{\rm XX-LA}$ associated with scattering of QW excitonic 
molecules by bulk LA-phonons. The calculations are done with 
Eq.\,(\ref{XX-LA}) for the X deformation potential $D_x = 8$\,eV 
(dashed line), 10\,eV (solid line), and 12\,eV (dotted line). }
\end{figure}

The values of the deformation potential $D_x$, published in literature, 
disperse in the band $7\,\mbox{eV} \leq D_x \leq 18$\,eV. In Fig.\,11 we 
plot $\Gamma^{\rm QW}_{\rm XX-LA} = \Gamma^{\rm QW}_{\rm XX-LA}(T)$ 
calculated by Eq.\,(\ref{XX-LA}) for $D_x =$ 8, 10, and 12\,eV. The 
deformation potential $D_x = 8$\,eV, which gives 
$\Gamma^{\rm QW}_{\rm XX-LA}(T\!=\!9\,\mbox{K}) \simeq 0.094\,$meV 
and is close to $D_x \simeq 9.6$\,eV reported for GaAs in 
Ref.\,[\onlinecite{Pollak68}], fits the temperature dependence 
$\Gamma^{\rm QW}_{\rm XX-LA} = \Gamma^{\rm QW}_{\rm XX-LA}(T)$ measured 
for the reference QW. In particular, 
$\Gamma^{\rm QW}_{\rm XX-LA}(T\!=\!10\,\mbox{K}) \simeq 0.1$\,meV is 
inferred from the total ${\tilde \Gamma}^{\rm QW}_{\rm XX} \simeq 
0.2\,$meV (see the inset of Fig.\,8). Thus from our FWM measurements of 
${\tilde \Gamma}^{\rm MC}_{\rm XX}$ at $T=9$\,K we conclude that the XX 
radiative width $\Gamma^{\rm MC}_{\rm XX} = 
{\tilde \Gamma}^{\rm MC}_{\rm XX} - \Gamma^{\rm QW}_{\rm XX-LA}$ is about 
$0.2-0.3$\,meV, i.e., is consistent with the values calculated within 
the bipolariton model (see Fig.\,10). 

In order to apply the bipolariton model [see Eq.\,(\ref{BPint})] to 
excitonic molecules in the reference single QW, one should take into 
account that the reference QW is sandwiched between a thick substrate 
and a cap layer of the thickness $L_{\rm cap} \simeq 499$\,nm. The 
evanescent light field associated with the QW polaritons is modified by 
the cap layer. Indeed, for the $- \epsilon_{\rm XX}^{(0)}/2$ energy 
detuning from the X resonance, one estimates that $\kappa \simeq 1.4 
\times 10^4\,\rm{cm}^{-1}$ so that $\exp(-\kappa L_{\rm cap}) \simeq 
0.5$ is not negligible. The estimate refers to two frequency-degenerate 
outgoing interface polaritons ($\hbar \omega = E_{\rm X} - 
\epsilon_{\rm XX}^{(0)}0/2$) created in the photon-assisted resonant 
dissociation of the QW molecule with ${\bf K}_{\|}=0$. At $z=L_{\rm cap}$ 
the initial evanescent field splits into two evanescent fields, 
``transmitted'' to air (or vacuum) and ``reflected'' back towards the QW. 
The first light field very effectively decays in the $z$-direction, with 
$\kappa_{\rm air} = \sqrt{ p^2_{\|} - (\omega/c)^2} \simeq 2.6 \times 
10^5\,\rm{cm}^{-1} \gg \kappa$. The ``reflected'' evanecsent light field 
makes at $z$=0 a destructive superposition with the initial evanecsent 
field, because the reflection coefficient of the top surface of the 
cap layer is $r_{\rm cap} = (\kappa - \kappa_{\rm air})/(\kappa + 
\kappa_{\rm air}) \simeq -0.9$. The destructive superposition stems from 
the $\pi$-jump of the phase of the ``reflected'' evanescsent field. Thus 
the effective oscillator strength relevant to the QW bipolariton wave 
Eq.\,(\ref{BPint}) is given by ${\tilde R}_{\rm X}^{\rm QW} = 
R_{\rm X}^{\rm QW}[1 + (r_{\rm cap}/2) \exp(- 2 \kappa L_{\rm cap})]^2$. 
For our reference structure with $\hbar^2 R_{\rm X}^{\rm QW} \simeq  
0.035\,{\rm eV}^2\AA$ we estimate $\hbar^2 {\tilde R}_{\rm X}^{\rm QW} 
\simeq 0.028\,{\rm eV}^2\AA$. In this case 
Eqs.\,(\ref{BPint})-(\ref{rada}) yield the total radiative width 
$\Gamma_{\rm XX}^{\rm QW}({\bf K}_{\|}$=0) $\simeq 0.126\,{\rm meV}$ 
(see Fig.\,3), the value which is very close to $\Gamma_{\rm XX}^{\rm QW} 
\simeq 0.1\,{\rm meV}$ obtained from the experimental data. 

Thus the bipolariton model, which attributes the XX radiative corrections 
mainly to the in-plane dissociation of molecules into outgoing 
interface/MC polaritons, reproduce quantitatively the XX radiative 
widths $\Gamma^{\rm MC}_{\rm XX}$ and $\Gamma^{\rm QW}_{\rm XX}$ 
estimated from the experimental data. The two main channels for the XX 
decay in microcavities, ``XX $\rightarrow$ 1$\lambda$-LB polariton + 
1$\lambda$-LB polariton'' and ``XX $\rightarrow$ 0$\lambda$-LB (or 
interface) polariton + 0$\lambda$-LB (or interface) polariton'' in 
comparison with the one leading decay route in single QWs, 
``XX $\rightarrow$ interface polariton + interface polariton'', explain 
qualitatively the factor two difference between 
$\Gamma^{\rm MC}_{\rm XX}$ and $\Gamma^{\rm QW}_{\rm XX}$. The 
XX-mediated optics of microcavities does require to include the 
``hidden'' 0$\lambda$-cavity (or interface, if the transverse optical 
confinement is relaxated for large ${\bf p}_{\|}$) polariton mode, which 
is invisible in standard optical experiments and, therefore, is usually 
neglected. Furthermore, with decreasing temperature $T \lesssim 10$\,K 
${\tilde \Gamma}^{\rm MC}_{\rm XX}$ and 
${\tilde \Gamma}^{\rm QW}_{\rm XX}$ effectively approach 
$\Gamma^{\rm MC}_{\rm XX}$ and $\Gamma^{\rm QW}_{\rm XX}$, respectively, 
so that the dephasing of the two-photon XX polarization in the 
microcavities and the reference QW occurs mainly through the optical 
decay of the molecules. Thus the $T_2=2T_1$ limit holds for the 
XX-mediated optics in our high-quality nanostructures and justifies the 
bipolariton model. The latter interprets the XX optical response in terms 
of resonant polariton-polariton scattering and requires nonperturbative 
treatment of both leading interactions, exciton-exciton Coulombic 
attraction and exciton-photon resonant coupling. Note that in our 
calculations with the exactly solvable bipolariton model only two 
control parameters of the theory, the input XX binding energy 
$\epsilon_{\rm XX}^{(0)}$ and the MC Rabi frequency 
$\Omega^{\rm MC}_{\rm 1\lambda}$ (or the X oscillator strength 
$R^{\rm QW}_{\rm X}$ for the reference QW), are taken from the 
experimental data. No fitting parameters are used in the numerical 
simulations. 

The relative motion of two optically-dressed constituent excitons of the 
bipolariton eigenstate (i.e., of the excitonic molecule) is affected by 
the exciton-photon interaction, according to the polariton dispersion 
law. The optically-induced change of the X energy occurs not only in 
the close vicinity of the resonant crossover between the initial photon 
and exciton dispersions, but in a rather broad band of ${\bf p}$ 
(or ${\bf p}_{\|}$). For example, in bulk semiconductors the effective 
mass associated with the upper polariton dispersion branch at $p=0$ is 
given by 
\begin{equation}
M^{\rm (3D)}_{\rm UB} \simeq { M_x \over 1 + 2 (\omega_{\ell t} / 
\omega_t) [(M_x c^2/\varepsilon_b)/(\hbar \omega_t)] } \, .  
\label{Mass3D}
\end{equation}
For bulk GaAs Eq.\,(\ref{Mass3D}) yields $M_{\rm eff} = 
M^{\rm (3D)}_{\rm UB}$ nearly by factor four less than the translational 
mass relevant to the pure excitonic dispersion, $M_x \simeq 0.7\,m_0$. 
From the microcavity dispersion Eq.\,(\ref{MC}) one estimates for $p_{\|} 
\rightarrow 0$ the effective masses associated with the 
1$\lambda$-eigenmode polariton dispersion branches: 
\begin{equation}
M^{\rm (MC)}_{\rm 1\lambda UB/LB}(\delta) \simeq { 2 E_{\rm X} \over 
(c^2/\varepsilon_b)[1 \pm \delta / (\hbar 
\Omega_{\rm 1\lambda}^{\rm MC})] } \, , 
\label{MassMC}
\end{equation}
where we assume that $|\delta| \lesssim \hbar 
\Omega_{\rm 1\lambda}^{\rm MC}$. In particular, for a zero-detuning 
GaAs-based microcavity Eq.\,(\ref{MassMC}) yields 
$M^{\rm (MC)}_{\rm 1\lambda UB}(\delta$=$0) = 
M^{\rm (MC)}_{\rm 1\lambda LB}(\delta$=$0) 
= 2E_{\rm X}/(c^2/\varepsilon_b) \simeq 0.7 \times 10^{-4}\,m_0$. In the 
meantime, at relatively large in-plane momenta $p_{\|} \sim 
p^{(1\lambda)}_{\|} < p_0$ the $1\lambda$-LB polariton energy smoothly 
approaches the exciton dispersion, i.e., $[E_{\rm X}(p_{\|}) - \hbar 
\omega^{\rm MC}_{\rm 1\lambda LB}(p_{\|})] \big|_{p_{\|} \sim 
p_{\|}^{(1\lambda)}} \rightarrow [\hbar 
(\Omega^{\rm MC}_{\rm 1\lambda})^2 \omega_t]/[2 
(c^2 p_{\|}^2/\varepsilon_b)] \propto 1/p_{\|}^2$, according to 
Eq.\,(\ref{MC}). While the above difference is rather small in absolute 
energy units, being compared with the in-plane kinetic energy of the 
exciton, $E_{\rm kin}^{\rm X} = \hbar^2 p_{\|}^2/2M_x$, it cannot be 
neglected. For example, the difference $E_{\rm X} - \hbar 
\omega^{\rm MC}_{\rm 1\lambda LB}$ becomes equal to $E_{\rm kin}^{\rm X}$ 
at $p_{\|} \simeq 1.35 \times 10^5\,\mbox{cm}^{-1}$. Note that for the 
above value of the in-plane wavevector $p_{\|}$ the photon component, 
associated with 1$\lambda$-LB polaritons, is negligible, i.e., 
$(u^{\rm MC}_{\rm 1\lambda LB})^2 \simeq 1 \gg 
(v^{\rm MC}_{\rm 1\lambda LB})^2$. Because it is a balance between the 
positive kinetic and negative interaction energies of the constituent 
excitons that gives rise to an excitonic molecule, the described 
optically-induced changes of the X effective mass at $p_{\|}=0$ and the 
nonparabolicity of the X dispersion at large $p_{\|}$ are responsible for 
the large XX radiative corrections in quasi-2D GaAs nanostructures.

\section{Conclusions}

In this paper we have studied, both theoretically and experimentally, 
the optical properties of QW excitonic molecules in semiconductor 
(GaAs) microcavities. We attribute the main channel of the XX optical 
decay to the resonant dissociation of MC molecules into outgoing MC 
polaritons, so that the XX-mediated optical signal we detect is due to 
the resonant radiative escape of the secondary MC polaritons through the 
DBRs. The bipolariton model has been adapted to construct the XX 
wavefunction ${\tilde \Psi}_{\rm XX}$ in terms of two (1$\lambda$-UB, 
1$\lambda$-LB and 0$\lambda$-LB) MC polaritons quasi-bound via Coulombic 
attraction of their exciton components. The MC bipolariton wave 
equation gives the radiative corrections to the XX state in 
microcavities. The following conclusions summarize our results. 

(i) The radiative corrections to the excitonic molecule state in 
GaAs-based microcavities, the XX Lamb shift $\Delta^{\rm MC}_{\rm XX}$ 
and the XX radiative width $\Gamma^{\rm MC}_{\rm XX}$, are large (about 
$0.15-0.30$ of the XX binding energy $\epsilon_{\rm XX}^{(0)}$) and 
definitely cannot be neglected. 

(ii) While usually the QW exciton -- mediated optics of semiconductor 
microcavities is formulated in terms of two 1$\lambda$-mode polariton 
dispersion branches only (1$\lambda$-UB and 1$\lambda$-LB, according to 
the terminology used in our paper), we emphasize the importance of the 
0$\lambda$-mode lower-branch polariton dispersion: The Coulombic 
interaction of the constituent excitons, which is responsible for the 
XX state, does couple intrinsically three relevant MC polariton branches, 
(1$\lambda$-UB, 1$\lambda$-LB, and 0$\lambda$-LB). Furthermore, the XX 
decay path ``XX $\rightarrow$ 0$\lambda$-LB polariton + 0$\lambda$-LB 
polariton'' is comparable in efficiency with the optical decay into 
1$\lambda$-LB polariton modes, i.e., ``XX $\rightarrow$ 1$\lambda$-LB 
polariton + 1$\lambda$-LB polariton''. Due do the relaxation of the DBR 
optical confinement for in-plane wavevectors $p_{\|} \sim p_0 = \omega_t 
\sqrt{\varepsilon_b}/c$, with increasing  $p_{\|}$ the 0$\lambda$-LB 
evolves towards the interface polariton dispersion associated with QW 
excitons. However, the short-wavelength LB polaritons with $p_{\|} \sim 
p_0$ always contribute to the  XX-mediated optics of microcavities. 

(iii) The zero-temperature extrapolation of the experimentally found XX 
dephasing width ${\tilde \Gamma}_{\rm XX}^{\rm MC}(T$=9\,K) yields 
$\Gamma_{\rm XX}^{\rm MC}(T$=0\,K) $\simeq 0.2-0.3$\,meV and is in a 
quantitative agreement with the result of the exactly solvable 
bipolariton model, $\Gamma_{\rm XX}^{\rm MC} \simeq 0.20-0.22$\,meV. 
From the analysis of the experimental data we conclude that the 
bipolariton model of MC excitonic molecules, which requires $T_2 \simeq 
2T_1$ limit, is valid for our high-quality GaAs-based nanostructures 
at $T \lesssim 10$\,K. For the reference GaAs QW without the DBR 
transverse optical confinement we find $\Gamma_{\rm XX}^{\rm QW} = 
{\tilde \Gamma}_{\rm XX}^{\rm QW}(T$=0\,K) $\simeq 0.1$\,meV. The latter 
value is also quantitatively consistent with that calculated by solving 
the QW bipolariton wave equation, $\Gamma_{\rm XX}^{\rm QW} = 0.126$\,meV. 
The nearly factor two difference between $\Gamma_{\rm XX}^{\rm QW}$ and 
$\Gamma_{\rm XX}^{\rm MC}$ clearly demonstrates the existence of the 
additional decay channel for a quasi-2D excitonic molecule in 
microcavities [``XX $\rightarrow$ interface polariton + interface 
polariton'' in MC-free single QWs versus ``XX $\rightarrow$ 0$\lambda$-LB 
(or interface) polariton + 0$\lambda$-LB (or interface) polariton'' and 
``XX $\rightarrow$ 1$\lambda$-LB polariton + 1$\lambda$-LB polariton'' 
for MC-embedded QW molecules]. 

(iv) The critical van Hove points, $M_1(\delta\!=\!\delta_1)$ and 
$M_2(\delta\!=\!\delta_2)$, in the JDPS of the resonant optical channel 
``XX (${\bf K}_{\|}$=0) $\leftrightarrow$ two 1$\lambda$-mode 
MC polaritons'' can allow us to find accurately the molecule binding 
energy $\epsilon^{\rm MC}_{\rm XX}$ and the MC Rabi frequency 
$\Omega_{\rm 1\lambda}^{\rm MC}$. Thus, by using time-dependent MC 
detuning $\delta = \delta(t)$, we propose to develop high-precision 
modulation spectroscopies in order to detect the rapid changes of the 
XX radiative corrections at $\delta = \delta_{1,2}$ [spikes in the XX 
Lamb shift $\Delta^{\rm MC}_{\rm XX} = \Delta^{\rm MC}_{\rm XX}(\delta 
= \delta_{1,2})$ and jumps in the XX radiatve width 
$\Gamma^{\rm MC}_{\rm XX} = \Gamma^{\rm MC}_{\rm XX}(\delta = 
\delta_{1,2})$] and estimate $\epsilon^{\rm MC}_{\rm XX}$ and 
$\Omega_{\rm 1\lambda}^{\rm MC}$.

\section{Acknowledgments}

We appreciate valuable discussions with J.~R. Jensen and J.~M. Hvam. 
Support of this work by the DFG, EPSRC and EU RTN Project 
HPRN-CT-2002-00298 is gratefully acknowledged.

%

\end{document}